\documentclass[lettersize,journal]{IEEEtran}
\IEEEoverridecommandlockouts
\usepackage{cite}
\usepackage{amsmath,amssymb,amsfonts}
\usepackage{algorithm,algorithmic}
\usepackage{graphicx}
\usepackage{textcomp}
\usepackage{amsmath}  
\usepackage{amsfonts}  
\usepackage{amssymb}  
\usepackage{soul}
\usepackage{enumitem}
\usepackage{xcolor}
\usepackage{booktabs}
\usepackage{cuted}
\usepackage{svg}
\usepackage{tcolorbox}
\usepackage{bm}
\usepackage{array}
\usepackage{makecell}
\usepackage{amsmath}
\usepackage{lettrine}

\usepackage{balance}

\usepackage[letterpaper, right=0.622in, left=0.622in, top=0.73in, bottom=1.02in]{geometry}

\def\BibTeX{{\rm B\kern-.05em{\sc i\kern-.025em b}\kern-.08em
    T\kern-.1667em\lower.7ex\hbox{E}\kern-.125emX}}

    \makeatletter
\renewcommand{\maketag@@@}[1]{\hbox{\m@th\normalsize\normalfont#1}}%
\makeatother

\begin{document}

\title{Movable Antenna Aided Full-Duplex ISAC System with Self-Interference Mitigation}
\author{

\IEEEauthorblockN{Size Peng,~\IEEEmembership{Student~Member,~IEEE}, Yin Xu,~\IEEEmembership{Senior~Member,~IEEE}, Guanli Yi, Cixiao Zhang,~\IEEEmembership{Student~Member,~IEEE}, \\
Dazhi He,~\IEEEmembership{Senior~Member,~IEEE}, and Wenjun Zhang,~\IEEEmembership{Fellow,~IEEE}}




    \thanks{
    This paper is supported in part by National Natural Science Foundation of China Program(62422111). An earlier version of this paper was presented in part at the IEEE Wireless Communications and Networking Conference 2025 [DOI:10.48550/arXiv.2408.00413].
    \textit{(Corresponding author: Yin Xu).}
    
Size Peng, Yin Xu, Cixiao Zhang, Dazhi He and Wenjun Zhang are with the Cooperative Medianet Innovation Center, Shanghai Jiao Tong University, Shanghai 200240, China. Dazhi He is also affiliated with Pengcheng Laboratory, Shenzhen 518055, China (e-mail: \{sjtu2019psz, xuyin, cixiaozhang, hedazhi, zhangwenjun\}@sjtu.edu.cn).
 
 
 
 Guanli Yi is with No.722 Research Institute of CSSC, Wuhan 430200, China (e-mail: yiguanli@sina.com).

 }
}

\markboth{Journal of \LaTeX\ Class Files,~Vol.~18, No.~9, September~2020}%
{How to Use the IEEEtran \LaTeX \ Templates}

\maketitle

\begin{abstract}
 Movable antenna (MA) has shown significant potential for improving the performance of integrated sensing and communication (ISAC) systems.
In this paper, we model an MA-aided ISAC system operating in a communication full-duplex mono-static sensing framework. The self-interference channel is modeled as a function of the antenna position vectors under the near-field channel condition. We develop an optimization problem to maximize the weighted sum of downlink and uplink communication rates alongside the mutual information relevant to the sensing task.
To address this highly non-convex problem, we employ the fractional programming (FP) method and propose an alternating optimization (AO)-based algorithm that jointly optimizes the beamforming, user power allocation, and antenna positions at the transceivers.
Given the sensitivity of the AO-based algorithm to the initial antenna positions, a PSO-based algorithm is proposed to explore superior sub-optimal antenna positions within the feasible region. 
 Numerical results indicate that the proposed algorithms enable the MA system to effectively leverage the antenna position flexibility for accurate beamforming in a complex ISAC scenario. This enhances the system's self-interference cancellation (SIC) capabilities and markedly improves its overall performance and reliability compared to conventional fixed-position antenna designs.
\end{abstract}


\begin{IEEEkeywords}
Movable antenna (MA), integrated sensing and communication (ISAC), full-duplex mono-static system, joint transceivers optimization, particle swarm optimization (PSO).
\end{IEEEkeywords}


\section{Introduction}
\IEEEPARstart{T}{he}
 increasing demand for reliable sensing and efficient communication has sparked significant interest in Integrated Sensing and Communication (ISAC) technologies. ISAC aims to merge communication and sensing functions within a single system, utilizing the same frequency bands and hardware resources. This integration improves spectral resource utilization, reduces hardware costs, and simplifies system complexity, positioning ISAC as a highly promising and efficient approach for modern wireless networks. Recent studies have demonstrated that ISAC systems can significantly enhance spectral efficiency compared to conventional systems that treat communication and sensing as separate functionalities \cite{survey-isac2}. As wireless networks evolve toward 6G and beyond, ISAC is anticipated to play a pivotal role in addressing the increasing demands for ultra-high data rates, low latency, and enhanced connectivity \cite{roadto6G}.

In an ISAC system, it is crucial to effectively manage the trade-off between communication and sensing capabilities and attain superior performance in both communication and sensing tasks.
To tackle this issue, extensive research efforts have been devoted to exploring effective ISAC system designs. For example, the work in \cite{CF_ISAC_MIMO} investigates the trade-off between communication and sensing in a cell-free multiple-input multiple-output (MIMO) ISAC system.
In \cite{ISAC_MIMO_POWER}, a combination of a linear precoder and a pre-designed array beamformer is proposed to minimize transmit power. The study in \cite{IRS_ISAC} leverages intelligent reflecting surfaces (IRSs) to enhance sensing performance and system security. In \cite{NOMA_ISAC}, non-orthogonal multiple access (NOMA) is used to improve both sensing and communication performance compared to the orthogonal multiple-access system.

Sensing in ISAC systems can be classified into three primary types \cite{survey_sensing}, \cite{survey-isac}: mono-static, bi-static, and multi-static. In mono-static sensing, the sensing signal would be sent and received by the same base station (BS). While in bi-static and multi-static sensing, different base stations are involved in the transmission and reception of the sensing signal. Notably, the full-duplex (FD) mono-static ISAC, which means communication and sensing signals would transmit simultaneously at the BS, is practically appealing for automotive and Internet of Things (IoT) applications due to its seamless integration, cost-effectiveness, and efficient use of the spectrum \cite{ISAC-survey}. 

Nonetheless, self-interference (SI) occurs between transmit and receive antennas in a mono-static ISAC system. 
To attain high system performance, it is imperative to have a robust Self-Interference Cancellation (SIC) capability. A conventional way to achieve SIC is by physical methods \cite{8115161}. Recent research has indicated that, in contrast to physical isolation methods, active suppression of self-interference (SI) can be realized via Tx and Rx beamforming \cite{SIC_survey,shi_robust_2022}. However, due to the limited precision of beamforming, this approach still impacts the overall system performance.

Moreover, in a more practical scenario, communication operates in an FD mode, implying that both uplink and downlink transmissions occur concurrently. In such a context, the ISAC system must give additional consideration to uplink communication performance, which imposes more stringent requirements on the system. 
Additionally, the uplink signal and sensing signal are prone to mutual interference, which will significantly degrade the system performance if the beamforming isn't designed appropriately. 
Several studies have discussed the solution to this issue by optimizing the beamforming design.
In \cite{ISACbeamforming}, ISAC with FD communication is discussed, and a joint beamforming and user power allocation solution is proposed. 
\cite{10571789}, a robust energy efficiency maximization problem is discussed, and a robust energy efficient beamforming design is proposed.

However, all of the aforementioned ISAC-related studies employ a fixed-position antenna (FPA) system, which restricts their ability to achieve superior performance in downlink and uplink communication, as well as in sensing tasks.
The primary reason underlying this limitation is the insufficient exploitation of the spatial degrees of freedom (DoF) offered by multiple antennas. To overcome this limitation, a movable antenna (MA) system \cite{R1}, also known as a fluid antenna  system (FAS) \cite{FAS}, has been proposed. This innovative system can flexibly adjust antenna positions, thereby capturing the spatial variations of wireless channels to enhance communication and sensing performance \cite{zhumulti}. 
To date, MA has exhibited its potential to enhance the performance of communication systems in a variety of applications \cite{zhuMAsurvey}, such as secure communication \cite{securemovable},  rate splitting multiple access \cite{zhang2024sumratemaximizationmovable}, and index modulation \cite{guo2024fluidantennagroupingbasedindex}.

Several studies have investigated the advantage of MA.
In \cite{Modeling_MA}, the author developed a field-response model in far-field conditions and analyzed the maximum channel gain with single receive MA. Results show that the MA system can reap considerable performance gains over the conventional FPA system.
In \cite{Capacity_Characterization}, the author modeled an MA-enabled MIMO system and investigated the channel capacity in the low-SNR regime. An alternating optimization algorithm is proposed to maximize the capacity, and a convex relaxation technique is applied to adjust the position of both transmit and receive MAs to obtain a locally optimal solution.
In \cite{MU_UL_MA}, the author modeled an MA-aided uplink communication system with multiuser and proposed a two-loop iterative algorithm to maximize the minimum achievable rate. In the algorithm, the antenna position is updated by particle swarm optimization (PSO), which could obtain a sub-optimal result efficiently.
In \cite{nearfield_MA}, a near-field multi-user communication system is discussed, and a two-loop dynamic neighborhood pruning PSO algorithm is proposed to solve the problem.
In \cite{lyu2024flexible}, the author modeled a bi-static ISAC system with movable antennas. They proposed a search-based projected gradient ascent (GA) method to update the antenna positions.
In \cite{peng2024jointantennapositionbeamforming}, the author modeled a mono-static MA-ISAC system and proposed a coarse-to-fine-grained searching algorithm to optimize the antenna position. They also take the SI channel into consideration and mitigate the SI with movable antennas.
In \cite{RIS_MA_ISAC}, the author enhanced the physical layer security performance of the ISAC system by employing movable antennas and reconfigurable intelligent surface. They proposed a two-layer penalty-based algorithm to achieve a trade-off between the optimality and feasibility of the solution.

Although there have been some research in related areas, the exploration of integrating MA into an ISAC system, especially in the context of FD communication, mono-static sensing, and SIC, remains scarce. To fill this gap, this paper models a communication FD mono-static sensing ISAC system aided with MA and takes into account SI. We delve into methods of enhancing the weighted sum of communication and sensing performance. Our contributions are briefly summarized as follows.
\begin{itemize}
    \item  We model a communication FD mono-static sensing ISAC system aided with MA. In this system, the BS is equipped with movable transmit and receive antennas. It serves multiple uplink and downlink users and simultaneously senses one target amid the interference of clutters while SI exists between the transmit and receive antennas.
    Moreover, to strike a balance among the uplink communication, downlink communication, and sensing performance, we formulate an optimization problem, which aims to maximize the weighted sum by jointly optimizing transmit and receive beamforming matrices, the uplink user power, and the antenna positions.    
    \item  To tackle the non-convex nature of the optimization problem, we employ fractional programming and formulate several subproblems. We propose an alternating optimization (AO) algorithm to iteratively optimize the beamforming matrices, auxiliary variables, and update antenna positions with a GA method to get a local optimal solution.
    Given that the performance of the AO-based algorithm highly depends on the initial position, we further propose a PSO-based algorithm. This algorithm can efficiently search for sub-optimal antenna positions within the entire feasible region, thereby highlighting the advantage of the movable region.
    \item  We conduct numerical simulations with varying parameters and analyze the performance of the proposed algorithms.  
    The results indicate that, when using the proposed AO-based algorithm, the MA system could perform better than the FPA system under different parameter settings. 
    Moreover, when applying the proposed PSO-based algorithm, there is more significant improvement in the performance of the MA system, highlighting the algorithm's exceptional capacity to exploit the movable region.
    Additionally, the PSO-based algorithm can handle more complex channel scenarios and is more adaptable to an enlarged feasible region.
\end{itemize}

The rest of this paper is organized as follows. Section \ref{sec:2} describes the system model and introduces a weighted sum maximization problem.
In Section \ref{sec:3}, algorithms for jointly optimizing beamforming, user power allocation, and antenna position to solve the problem are proposed. 
In Section \ref{sec:4}, numerical results are presented to demonstrate the performance.
And a conclusion is given in Section \ref{sec:5}.

\textit{Notations:}  Scalar variables are denoted by italic letters, vectors are denoted by boldface small letters, and matrices are denoted by boldface capital letters. $\mathbf{X}(n)$, $\mathbf{X}^{T}$, $\mathbf{X}^{*}$, $\text{Tr}(\mathbf{X})$, $(\mathbf{X})^{-1}$,  $[\mathbf{X}]_{i,j}$,$||\mathbf{X}||$, $|x|$, $\mathbf{X}^H$, $\text{Re}(\mathbf{x})$ and $\nabla_y{\mathbf{X}}$ denote the $n$-th entry of $\mathbf{X}$, the transpose of $\mathbf{X}$, the conjugate of $\mathbf{X}$, the trace of $\mathbf{X}$, the inverse of $\mathbf{X}$, the entry in the $i$-th row and $j$-th column of the matrix $\mathbf{X}$,  the L-2 norm of $\mathbf{X}$,  the absolute value of $x$, the conjugate transpose of $\mathbf{X}$, the real part of $\mathbf{x}$ and the partial derivative of $\mathbf{X}$ with respect to $y$ respectively. 
$j$ denotes the imaginary unit.
$\mathbb{C}^{M \times N}$ is the set of complex matrices with $M$ rows and $N$ columns.

\section{System Model} \label{sec:2}
\begin{figure}[t]
    \centering
    \includegraphics[width=0.93\linewidth]{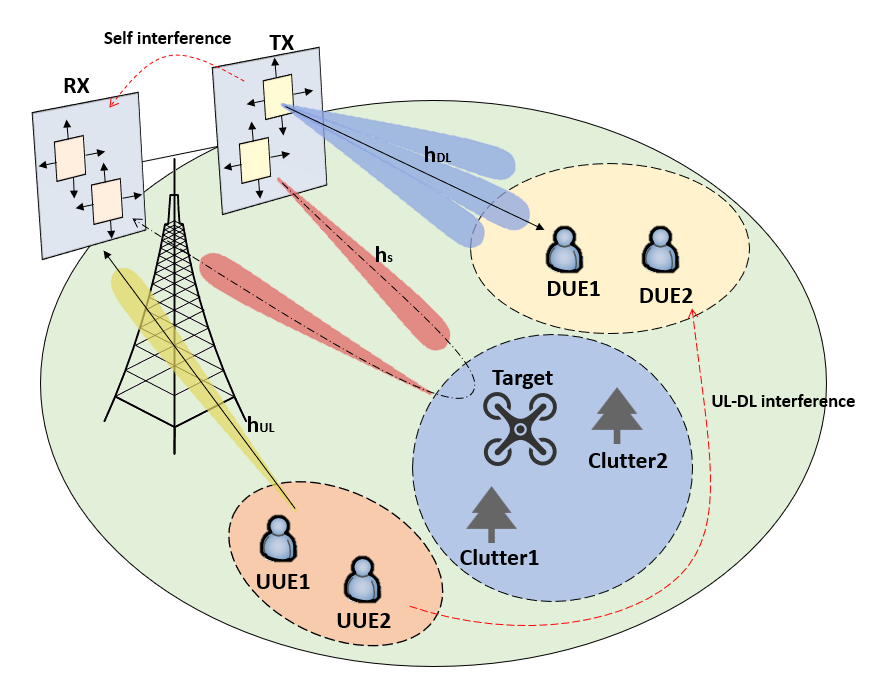}
    \caption{Illustration of the proposed ISAC system aided with MA.}
    \label{fig:system_model}
\end{figure}

In this paper, we consider a communication FD mono-static sensing BS equipped with $N_T$ transmit antennas and $N_R$ receive antennas simultaneously serving $K_{\text{UL}}$ uplink users and $K_{\text{DL}}$ downlink users while sensing for one target amid interference from $C$ clutters. The system setup is illustrated in Fig.\ref{fig:system_model}. The transmit and receive antennas are independently movable within distinct two-dimensional planes, each sharing an identical feasible region. Specifically, the antenna positions are confined to the horizontal coordinates $[X_{\text{min}}, X_{\text{max}}]$ and vertical coordinates $[Y_{\text{min}}, Y_{\text{max}}]$. Furthermore, self-interference exists between the transmit and receive antennas in this configuration.

\subsection{Channel Model}
Let $\mathbf{p}_t$ and $\mathbf{p}_r$ denote the positions of transmit and receive antennas, respectively. The position of the $m$-th transmit antenna is $\mathbf{p}_{t,m}=[x_{t,m},y_{t,m}]^T$, while the position of the $n$-th receive antenna is $\mathbf{p}_{r,n}=[x_{r,n},y_{r,n}]^T$. Let $\boldsymbol{\theta}_{\text{t}}$ and $\boldsymbol{\phi}_{\text{t}}$
denote the azimuth and elevation angles of departure (AoDs) for the downlink transmission, respectively. And let $\boldsymbol{\theta}_{\text{r}}$ and $\boldsymbol{\phi}_{\text{r}}$
denote the azimuth and elevation angles of arrival (AoAs) of the uplink transmission. The difference in transmission distance between the $i$-th transmit antenna and the origin within the feasible region for the $l$-th path of the $k$-th downlink user is given by
\begin{equation}
r_{t,k,l,i}=x_{t,i}\cos(\theta_{t,k,l})\sin(\phi_{t,k,l})+y_{t,i}\sin(\theta_{t,k,l}).
\end{equation}
Accordingly, the transmit steering vector for the $l$-th path of the $k$-th downlink user is expressed as
\begin{equation}
\mathbf{a}_{k,l}(\mathbf{p}_t) = \begin{bmatrix} e^{j\frac{2\pi }{\lambda} r_{t,k,l,1}},\cdots, e^{j\frac{2\pi }{\lambda} r_{t,k,l,N_T}} \end{bmatrix}^T\in \mathbb{C}^{N_T}.
\label{steering_vector}
\end{equation}
We assume that there are $L_p$ paths for the downlink channel. Accordingly, the channel of the $k$-th downlink user is modeled as
\begin{equation}
    \mathbf{h}_{\text{DL},k}(\mathbf{p}_t) = \sqrt{\frac{\eta_k}{L_p}} \sum\limits_{l=1}^{L_p} \rho_{\text{DL},k,l} \mathbf{a}_{k,l}(\mathbf{p}_t)\in \mathbb{C}^{N_T}.
\end{equation}
Similarly, for the receive antennas, the difference in transmission distance between the \(j\)-th receive antenna and the origin within the feasible region for the \(l\)-th path of the \(k\)-th uplink user is given by 
\begin{equation}
r_{r,k,l,j}=x_{r,j}\cos(\theta_{r,k,l})\sin(\phi_{r,k,l})+y_{r,j}\sin(\theta_{r,k,l}).
\end{equation}
The receive steering vector for the $l$-th path of the $k$-th uplink user is
\begin{equation}
    \mathbf{b}_{k,l}(\mathbf{p}_r) = \begin{bmatrix} e^{j\frac{2\pi }{\lambda} r_{r,k,l,1}},\cdots,e^{j\frac{2\pi }{\lambda} r_{r,k,l,N_R}} \end{bmatrix}^T\in \mathbb{C}^{N_R}.
\end{equation}
We assume that the uplink channel also consists of \(L_p\) propagation paths. Then the channel of the $k$-th uplink user can be denoted as
\begin{equation}
    \mathbf{h}_{\text{UL},k}(\mathbf{p}_r) = \sqrt{\frac{\eta_k}{L_p}} \sum\limits_{l=1}^{L_p} \rho_{\text{UL},k,l} \mathbf{b}_{k,l}^H(\mathbf{p}_r)\in \mathbb{C}^{1 \times N_R},
\end{equation}
where $\rho_{\text{DL},k,l}$, $\rho_{\text{UL},k,l}$, $\eta_k$ denote channel gain of downlink communication, channel gain of uplink communication, and free-space path loss, respectively.
Let $g_{i,j}$ represent the interference channel for $i$-th uplink user to $j$-th downlink user, which is 
\begin{equation}
    g_{i,j}=\sqrt{\eta_{i,j}}e^{-j \frac{2\pi}{\lambda} r_{i,j} },
\end{equation}
where \(r_{i,j}\) denotes the distance between the \(i\)-th uplink user and the \(j\)-th downlink user, and \(\eta_{i,j}\) represents the corresponding large-scale channel gain.

For sensing target and clutters, we assume that the channels have only one path, thus the steering vectors for sensing target and $c$-th clutter of transmit antennas are
\begin{equation}
\mathbf{a}_s(\mathbf{p}_t) = \begin{bmatrix} e^{j\frac{2\pi }{\lambda} r_{t,s,1}},\cdots, e^{j\frac{2\pi}{\lambda} r_{t,s,N_T}} \end{bmatrix}^T\in \mathbb{C}^{N_T},
\label{steering_vector}
\end{equation}
\begin{equation}
\mathbf{a}_c(\mathbf{p}_t) = \begin{bmatrix} e^{j\frac{2\pi }{\lambda} r_{t,c,1}},\cdots, e^{j\frac{2\pi }{\lambda} r_{t,c,N_T}} \end{bmatrix}^T\in \mathbb{C}^{N_T}.
\label{steering_vector}
\end{equation}
Similarly, the receive steering vectors are given by
\begin{equation}
\mathbf{b}_{s}(\mathbf{p}_r) = \begin{bmatrix} e^{j\frac{2\pi }{\lambda} r_{r,s,1}},\cdots,e^{j\frac{2\pi }{\lambda} r_{r,s,N_R}} \end{bmatrix}^T\in \mathbb{C}^{N_R},
\end{equation}
\begin{equation}
\mathbf{b}_{c}(\mathbf{p}_r) = \begin{bmatrix} e^{j\frac{2\pi }{\lambda} r_{r,c,1}},\cdots,e^{j\frac{2\pi }{\lambda} r_{r,c,N_R}} \end{bmatrix}^T\in \mathbb{C}^{N_R}.
\end{equation}
Here, $r_{t,s,i}$, $r_{r,s,j}$, $r_{t,c,i}$, and $r_{r,c,j}$ denote the difference in transmission distance between the origin within the feasible region and transmit and receive antennas. 
Specifically $r_{t,s,i}$ and $r_{r,s,j}$ correspond to the sensing target
whereas $r_{t,c,i}$ and $r_{r,c,j}$ pertain to the $c$-th clutter.
We could derive $r_{t,s,i}$ and $r_{t,c,i}$ as
 \begin{equation}
r_{t,s,i}=x_{t,i}\cos(\theta_{t,s})\sin(\phi_{t,s})+y_{t,i}\sin(\theta_{t,s}),
\end{equation}
\begin{equation}
r_{t,c,i}=x_{t,i}\cos(\theta_{t,c})\sin(\phi_{t,c})+y_{t,i}\sin(\theta_{t,c}),
\end{equation}
where $\theta_{t,s}$ and $\phi_{t,s}$ are the azimuth and elevation AODs of sensing target, $\theta_{t,c}$ and $\phi_{t,c}$ are the azimuth and elevation AODs of $c$-th clutter. Similarly we could derive $r_{r,s,j}$ and $r_{r,c,j}$ with the azimuth and elevation AOAs of sensing target and $c$-th clutter $\theta_{r,s}$, $\phi_{r,s}$, $\theta_{r,c}$ and $\phi_{r,c}$.

The channel of sensing target and clutters are
\begin{equation}
    \mathbf{h}_s(\mathbf{p}_t,\mathbf{p}_r)=\sqrt{\eta_s} \alpha_{s} \mathbf{a}_{s}(\mathbf{p}_t)\mathbf{b}_{s}^H(\mathbf{p}_r)\in \mathbb{C}^{N_T\times N_R},
\end{equation}
\begin{equation}
    \mathbf{h}_c(\mathbf{p}_t,\mathbf{p}_r)=\sqrt{\eta_c} \alpha_{c} \mathbf{a}_{c}(\mathbf{p}_t)\mathbf{b}_{c}^H(\mathbf{p}_r)\in \mathbb{C}^{N_T\times N_R},
\end{equation}
where the complex coefficients $\alpha_{s}$ and $\alpha_{c}$ represent the radar cross section (RCS) of the sensing target and the $c$-th clutter, respectively. Likewise, $\eta_s$ and $\eta_c$ denote the free-space path losses along the paths of the target and the $c$-th clutter, respectively. The path loss coefficient of communication and sensing follows the far-field modeling, which has $\eta=\big[\frac{\sqrt{G_l\lambda}}{4\pi d}\big]^2$, $d$ is the distance that the signal is transmitted, and $G_l$ is the free-space fading factor.

The channel of self-interference is
    \begin{equation}
[\mathbf{H}_{\text{SI}}]_{i,j}=\left(\sqrt{[\boldsymbol{\eta}_{\text{SI}}]_{i,j}}\ e^{-j \frac{2\pi}{\lambda} r_{ \text{SI},x_i,y_j}}\right),
\end{equation}
 where $\boldsymbol{\eta}_{\text{SI}}$ follows a near-field modeling \cite{near_field_modeling} 
 \begin{equation}
    \small
     [\boldsymbol{\eta}_{\text{SI}}]_{i,j} \!=\! \frac{G_l}{4}\bigg[\big(\frac{\lambda}{2\pi r_{\text{SI},x_i,y_j}}\big)^2\!-\!\big(\frac{\lambda}{2\pi r_{\text{SI},x_i,y_j}}\big)^4 \!+ \! \big(\frac{\lambda}{2\pi r_{\text{SI},x_i,y_j}}\big)^6\bigg].
 \end{equation}
The distance of self-interference channel is
\begin{equation}
    \mathbf{r}_{\text{SI},i,j}=\sqrt{(x_{t,j}-x_{r,i}+d_{\text{SI}})^2+(y_{t,i}-y_{r,j})^2},
\end{equation}
where $d_{\text{SI}}$ is the distance between the transmit antenna region and the receive antenna region.

\subsection{Signal Model}

Let $\mathbf{s_{\text{DL}}}=[s_{\text{DL},1},s_{\text{DL},2},\cdots,s_{\text{DL},K_{\text{DL}}}]^T$, $\mathbb{E}\{ \mathbf{s}_{\text{DL}} \mathbf{s}_{\text{DL}}^H \} = \mathbf{I}$, denote the communication signal of downlink users.
$\mathbf{s}_{\text{UL}}=[s_{\text{UL},1},s_{\text{UL},2},\cdots,s_{\text{UL},K_{\text{UL}}}]^T$, $\mathbb{E}\{ \mathbf{s}_{\text{UL}} \mathbf{s}_{\text{UL}}^H \} = \mathbf{I}$, denote the uplink communication signal.

The following are the beamforming matrices of transmit antennas, 
receive matrices for sensing,
receive matrices for uplink communication,
and uplink user precoding coefficients.
\begin{equation}
    \mathbf{F}=[\mathbf{f}_1,\mathbf{f}_2,...,\mathbf{f}_{K_{\text{DL}}}]\in \mathbb{C}^{N_T \times K_{\text{DL}}},
\end{equation}
\begin{equation}
    \mathbf{w}_s=[w_{s,1},w_{s,2},...,w_{s,N_R}]^T\in \mathbb{C}^{N_R},
\end{equation}
\begin{equation}
    \mathbf{w}_r=[\mathbf{w}_{r,1},\mathbf{w}_{r,2},...,\mathbf{w}_{r,{K_{\text{UL}}}}]\in \mathbb{C}^{N_R \times K_{\text{UL}}},
\end{equation}
\begin{equation}
    \mathbf{f}_{\text{UL}}=[f_{\text{UL},1},f_{\text{UL},2},\cdots,f_{\text{UL},K_{\text{UL}}}]^T\in \mathbb{C}^{K_{\text{UL}}}.
\end{equation}
The signal that the $k$-th downlink user receives is 
\begin{equation}
    \small
    r_k \! =  \!  \mathbf{h}_{\text{DL},k}^H(\mathbf{p}_t)\mathbf{f}_k s_{\text{DL},k} + \mathbf{h}_{\text{DL},k}^H\sum\limits_{j=1, j\neq k}^{K_{\text{DL}}} \mathbf{f}_j s_{\text{DL},j} + \sum\limits_{j=1}^{K_{\text{UL}}} g_{j,k} s_{\text{UL},j} + n_k. 
\end{equation}
The received signal is formulated with downlink signal for the $k$-th  user, downlink signal for other users, uplink signal interference, and  receive noise which follows $n_k \sim \mathcal{C}\mathcal{N}(0,\sigma_c^2)$. Therefore, the signal-to-interference-plus-noise ratio (SINR) for the $k$-th downlink user can be denoted as
\begin{equation}
\text{SINR}_{\text{DL},k} = \frac{|\mathbf{h}_{\text{DL},k}^H(\mathbf{p}_t)\mathbf{f}_k|^2}{\sum\limits_{j=1, j \neq k}^{K_{\text{DL}}} |\mathbf{h}_{\text{DL},k}^H(\mathbf{p}_t)\mathbf{f}_j|^2 + \sum\limits_{j=1}^{K_{\text{UL}}} |g_{j,k}|^2+\sigma_k^2}.
\end{equation}
The downlink communication rate for the $k$-th downlink user is
\begin{equation}
    R_{\text{DL},k}=\log_2(1+\text{SINR}_{\text{DL},k}).
\end{equation}
The received signal of the BS is
\begin{equation}
\begin{aligned}
       &\mathbf{r}_s =  \sqrt{\eta_s}\alpha_s \mathbf{b}_s(\mathbf{p}_r) \mathbf{a}_s^H(\mathbf{p}_t) \mathbf{F} \mathbf{s}_{\text{DL}} 
         + \sum\limits_{j=1}^{K_{\text{UL}}}  \mathbf{h}_{\text{UL},j}^H(\mathbf{p}_r) f_{\text{UL},j} s_{\text{UL},j}\\
        &+\mathbf{H}_{\text{SI}}^H(\mathbf{p}_t, \mathbf{p}_r) \mathbf{F} \mathbf{s}_{\text{DL}} + \sum_{c=1}^C \sqrt{\eta_c}\alpha_c\mathbf{b}_c(\mathbf{p}_r) \mathbf{a}_c^H(\mathbf{p}_t) \mathbf{F} \mathbf{s}_{\text{DL}}+\mathbf{n}_s.
\end{aligned}
\end{equation}
The received signal consists of the reflected sensing signal reflected by the target and clutters, self-interference from transmit antennas, uplink communication signal, and noise which follows $\mathbf{n}_s \sim \mathcal{C}\mathcal{N}(0,\sigma_s^2\mathbf{I})$.
In order to streamline the analysis, we define  
\begin{equation}
    \begin{aligned}
        &C_s=\sum\nolimits_{c=1}^{C} ||\sqrt{\eta_c}\alpha_c\mathbf{w}_s^H \mathbf{b}_c(\mathbf{p}_r)\mathbf{a}_c^H(\mathbf{p}_t) \mathbf{F}||^2,\\
        &C_{r,k}=\sum\nolimits_{c=1}^{C} ||\sqrt{\eta_c}\alpha_c\mathbf{w}_{r,k}^H \mathbf{b}_c(\mathbf{p}_r)\mathbf{a}_c^H(\mathbf{p}_t) \mathbf{F}||^2,\\
        &S_s=||\sqrt{\eta_s}\alpha_s\mathbf{w}_s^H \mathbf{b}_s(\mathbf{p}_r)\mathbf{a}_s^H(\mathbf{p}_t) \mathbf{F}||^2,\\
        &S_{r,k}=||\sqrt{\eta_s}\alpha_s\mathbf{w}_{r,k}^H \mathbf{b}_s(\mathbf{p}_r)\mathbf{a}_s^H(\mathbf{p}_t) \mathbf{F}||^2,\\
        &SI_s=||\mathbf{w}_s^H \mathbf{H}_{\text{SI}}^H(\mathbf{p}_t,\mathbf{p} _r)\mathbf{F}||^2, \notag \\
        &SI_{r,k}=||\mathbf{w}_{r,k}^H \mathbf{H}_{\text{SI}}^H(\mathbf{p}_t,\mathbf{p} _r)\mathbf{F}||^2, \notag
    \end{aligned}
\end{equation}
to denote the power of sensing signal from clutters and target received with sensing receive matrices and uplink communication receive matrices respectively.
As well as the power of self-interference signal received by the same respective matrices.

Thus the uplink SINR for the $k$-th uplink user can be derived as
\begin{equation}
\scalebox{0.88}{$
    \text{SINR}_{\text{UL},k} = \frac{| \mathbf{w}^H_{r,k} \mathbf{h}_{\text{UL},k}^H (\mathbf{p}_r) \mathbf{f}_{\text{UL},k} |^2 }
    {C_{r,k} + S_{r,k} + SI_{r,k} + \sum\limits_{j=1,j\neq k }^{K_{\text{UL}}} | \mathbf{w}^H_{r,k} \mathbf{h}_{\text{UL},j}^H(\mathbf{p}_r) f_{\text{UL},j} |^2 + ||\mathbf{w}_{r,k}||^2\sigma_s^2}.
    $}
\end{equation}
The uplink communication rate for the $k$-th uplink user is
\begin{equation}
    R_{\text{UL},k}=\log_2(1+\text{SINR}_{\text{UL},k}).
\end{equation}
The signal-to-clutter-plus-noise ratio (SCNR) at the BS is expressed as
\begin{equation}
\small
\text{SCNR}\! = \! \frac{S_s}{C_s + SI_s + \sum\limits_{j=1}^{K_{\text{UL}}} | \mathbf{w}^H_{s} \mathbf{h}_{\text{UL},j}^H(\mathbf{p}_r) f_{\text{UL},j} |^2 +||\mathbf{w}_s||^2\sigma_s^2} .
\end{equation}
The sensing mutual information (MI) can be expressed as
\begin{equation}
    R_{s}=\log_2(1+\text{SCNR}).
\end{equation}
\subsection{Problem Formulation}
To balance the uplink communication, downlink communication, and sensing performance, we aim to maximize the weighted sum of uplink communication rate, downlink communication rate, and sensing MI as
\begin{equation}
\begin{aligned}
    \mathcal{G}(\mathbf{F}, \mathbf{p}_t, \mathbf{p}_r, \mathbf{w}_r,\mathbf{w}_s,\mathbf{f}_{\text{UL}})& = \varpi_s R_s + \varpi_{c,\text{DL}} \sum_{k=1}^{K_{\text{DL}}} R_{\text{DL},k}   \\
    & +\varpi_{c,\text{UL}} \sum_{k=1}^{K_{\text{UL}}} R_{\text{UL},k}.
    \label{eq:orig_G}
\end{aligned}
\end{equation}
The optimization problem is formulated as
\begin{subequations}
\begin{align}
(\text{P1}) \max_{\mathbf{F}, \mathbf{p}_t, \mathbf{p}_r, \mathbf{w}_r,\mathbf{w}_s,\mathbf{f}_{\text{UL}}}  \mathcal{G}(\mathbf{F}, \mathbf{p}_t, &\mathbf{p}_r,\mathbf{w}_r,\mathbf{w}_s,\mathbf{f}_{\text{UL}})  \label{eq:originalG} \\
\quad \text{s.t.} \quad  \text{Tr}(\mathbf{F}^H \mathbf{F})& \leq P_{\text{DL}},  \label{eq:F}  \\
\text{Tr}(\mathbf{f}_{\text{UL}}^H\mathbf{f}_{\text{UL}}) &\leq P_{\text{UL}}, \label{eq:UL_P} \\
\quad X_{\min} \leq x_{t,i} \leq X_{\max}, Y_{\min} &\leq y_{t,i} \! \leq \! Y_{\max},  \forall i, \label{t_range_constraint} \\
\quad X_{\min} \leq x_{r,j} \leq X_{\max}, Y_{\min} &\leq y_{r,j} \!\leq \! Y_{\max},  \forall j, \label{r_range_constraint} \\
||p_{t,i} - p_{t,\hat{i}}|| \geq D_0, ||p_{r,j} - p_{r,\hat{j}}||& \geq D_0, i\neq\hat{i}, j\neq\hat{j} \label{distance_constraint}.
\end{align}
\end{subequations}
Here, $P_{\text{DL}}$ denotes the maximum transmit power of the BS, while $P_{\text{UL}}$ represents the total maximum transmit power of all uplink users. The parameter $D_0$ specifies the minimum allowable separation distance between any pair of antennas to mitigate coupling effects. Constraint~(\ref{eq:F}) imposes a limit on the total downlink transmit power, whereas constraint~(\ref{eq:UL_P}) restricts the total uplink transmission power from the uplink users. Constraints~(\ref{t_range_constraint}), (\ref{r_range_constraint}), and (\ref{distance_constraint}) ensure that the transmit and receive antennas operate within their designated feasible regions and maintain sufficient separation to satisfy hardware limitations. The weighting coefficients are constrained by $\varpi_{c,\text{DL}} + \varpi_{c,\text{UL}} + \varpi_{s} = 1$.

\section{Proposed Solution} \label{sec:3}
It is challenging to solve (P1) directly since the optimization function (\ref{eq:originalG})
is non-convex w.r.t. $\mathbf{F}, \mathbf{p}_t, \mathbf{p}_r, \mathbf{w}_r,\mathbf{w}_s,\mathbf{f}_{\text{UL}}$. To address this problem, we employ the fractional programming (FP)\cite{FPmethod}. We first introduce auxiliary variables $\boldsymbol{\mu}=[\mu_1,\cdots,\mu_{K_{\text{DL}}+K_{\text{UL}}+1}]^{T}$ and formulate the Lagrangian dual problem of (\ref{eq:orig_G}) as (\ref{equ:larg_t}). Then we use quadratic transform and introduce auxiliary variables $\boldsymbol{\xi}^{c,\text{DL}}=[\xi^{c,\text{DL}}_{1},\cdots,\xi^{c,\text{DL}}_{K_{\text{DL}}}]^{T}$, $\boldsymbol{\xi}^{c,\text{UL}}=[\xi^{c,\text{UL}}_1,\cdots,\xi^{c,\text{UL}}_{K_{\text{UL}}}]^{T}$ and $\boldsymbol{\xi}^s=[\xi^s_{1},\cdots,\xi^s_{K_{\text{DL}}}]^{T}$  to further transform (\ref{equ:larg_t}) into (\ref{eq:convexG}).
\begin{figure*}[t]
\small
\setlength\abovedisplayskip{3pt}
\setlength\belowdisplayskip{3pt}
\begin{align}
            &{\mathcal{G}}_l(\mathbf{F}, \mathbf{p}_t, \mathbf{p}_r, \mathbf{w}_r,\mathbf{w}_s , \mathbf{f}_{\text{UL}}, \boldsymbol{\mu})=\varpi_{c,\text{DL}} \sum\limits_{k=1}^{K_{\text{DL}}} \log(1+\mu_k) +\varpi_{c,\text{UL}} \sum\limits_{k=1}^{K_{\text{UL}}} \log(1+\mu_{K_{\text{DL}}+k}) +\varpi_s \log(1+\mu_{K_{\text{DL}}+K_{\text{UL}}+1})  -\varpi_{c,\text{DL}} \sum\limits_{k=1}^{K_{\text{DL}}} \mu_k  \notag \\
    &- \varpi_{c,\text{UL}} \sum\limits_{k=1}^{K_{\text{UL}}} \mu_{K_{\text{DL}}+k} - \varpi_s  \mu_{K_{\text{DL}}+K_{\text{UL}}+1} +    \varpi_s  \frac{(1+\mu_{K_{\text{DL}}+K_{\text{UL}}+1})S_s}{C_s + S_s +SI_s + \sum\limits_{j=1}^{K_{\text{UL}}} | \mathbf{w}^H_s \mathbf{h}_{\text{UL},j}^H(\mathbf{p}_r) f_{\text{UL},j} |^2 +||\mathbf{w}_{s}||^2\sigma_s^2}    \notag \\
    &       +  \varpi_{c,\text{DL}} \sum\limits_{k=1}^{K_{\text{DL}}} \frac{(1+\mu_k) |\mathbf{h}_{\text{DL},k}^H(\mathbf{p}_t)\mathbf{f}_k|^2  }{  \sum\limits_{j=1}^{K_{\text{DL}}} |\mathbf{h}_{\text{DL},k}^H(\mathbf{p}_t)\mathbf{f}_j|^2 + \sum\limits_{j=1}^{K_{\text{UL}}} |g_{j,k}|^2+\sigma_k^2 
 }   + \varpi_{c,\text{UL}} \sum_{k=1}^{K_{\text{UL}}} \frac{(1+\mu_{K_{\text{DL}}+k})  | \mathbf{w}^H_{r,k} \mathbf{h}_{\text{UL},k}^H(\mathbf{p}_r) f_{\text{UL},k} |^2  }{C_{r,k} + S_{r,k} +SI_{r,k} + \sum\limits_{j=1}^{K_{\text{UL}}} | \mathbf{w}_{r,k}^H \mathbf{h}_{\text{UL},j}^H(\mathbf{p}_r) f_{\text{UL},j} |^2 + ||\mathbf{w}_{r,k}||^2 \sigma_s^2}.    \label{equ:larg_t} \\
 &\hat{\mathcal{G}}(\mathbf{F}, \mathbf{p}_t, \mathbf{p}_r, \mathbf{w}_r, \mathbf{w}_s ,\mathbf{f}_{\text{UL}}, \boldsymbol{\mu},\boldsymbol{\xi}^{c,\text{DL}},\boldsymbol{\xi}^{c,\text{UL}},\boldsymbol{\xi}^s) = \varpi_{c,\text{DL}} \sum\limits_{k=1}^{K_{\text{DL}}} \log(1+\mu_k) +\varpi_{c,\text{UL}} \sum\limits_{k=1}^{K_{\text{UL}}} \log(1+\mu_{K_{\text{DL}}+k}) +\varpi_s \log(1+\mu_{K_{\text{DL}}+K_{\text{UL}}+1})  \notag  \\
 & +\! \varpi_s \Bigg\{ 2\sqrt{1+\mu_{K_{\text{DL}}+K_{\text{UL}}+1}} \text{Re}\left(\sqrt{\eta_s}\alpha_s \mathbf{w}^H_s \mathbf{b}_s(\mathbf{p}_r) \mathbf{a}_s^H(\mathbf{p}_t) \mathbf{F} \boldsymbol{\xi}^s\right) \! - \!  ||\boldsymbol{\xi}^s||^2 \! \left[ \!   C_s + S_s +SI_s + \sum\limits_{j=1}^{K_{\text{UL}}} | \mathbf{w}^H_s \mathbf{h}_{\text{UL},j}^H(\mathbf{p}_r) f_{\text{UL},j} |^2 + ||\mathbf{w}_s||^2 \sigma_s^2   \!  \right]  \! \Bigg\} \notag \\
        &  + \varpi_{c,\text{DL}} \sum\limits_{k=1}^{K_{\text{DL}}} \Bigg\{2\sqrt{1+\mu_k} \text{Re}\left(\xi^{c,\text{DL}}_k \mathbf{h}_{\text{DL},k}^H(\mathbf{p}_t) \mathbf{f}_k\right) - |\xi^{c,\text{DL}}_k|^2 \left[\sum_{j=1}^{K_{\text{DL}}} |\mathbf{h}_{\text{DL},k}^H(\mathbf{p}_t) \mathbf{f}_j|^2 + \sum\limits_{j=1}^{K_{\text{UL}}} |g_{j,k}|^2 +\sigma_k^2\right]\Bigg\} \notag \\
        &+\! \varpi_{c,\text{UL}} \! \sum_{k=1}^{K_{\text{UL}}}\! \Bigg\{ \! 2 \sqrt{1+\mu_{K_{\text{DL}}+k}} \text{Re}\left( \xi^{c,\text{UL}}_k f_{\text{UL},k}  \mathbf{w}^H_{r,k} \mathbf{h}_{\text{UL},k}^H(\mathbf{p}_r)  \right)  \! - \!   |\xi^{c,\text{UL}}_k|^2 \! \left[ \!  C_{r,k} + S_{r,k} +SI_{r,k} + \sum\limits_{j=1}^{K_{\text{UL}}} | \mathbf{w}_{r,k}^H \mathbf{h}_{\text{UL},j}^H(\mathbf{p}_r) f_{\text{UL},j} |^2 + ||\mathbf{w}_{r,k}||^2\sigma_s^2 \! \right] \! \Bigg\} \notag \\
        & -\varpi_{c,\text{DL}} \sum\limits_{k=1}^{K_{\text{DL}}} \mu_k  -\varpi_{c,\text{UL}} \sum\limits_{k=1}^{K_{\text{UL}}} \mu_{K_{\text{DL}}+k} -\varpi_{s} \mu_{K_{\text{DL}}+K_{\text{UL}}+1} .
        \label{eq:convexG}
\end{align}
\hrulefill
\end{figure*}
We formulate several subproblems and present an AO-based algorithm for optimization.
In the following subsections, we solve these subproblems respectively.

\subsection{Beamforming Optimization}
To optimize transmit beamforming matrix $\mathbf{F}$, we need to solve subprolem as
\begin{equation*}
    \begin{aligned}
        (\text{SP.1}) \max_{\mathbf{F}}\hat{\mathcal{G}}(\mathbf{F}|\mathbf{p}_t, \mathbf{p}_r, \mathbf{w}_r, \mathbf{w}_s, \mathbf{f}_{\text{UL}}, \boldsymbol{\mu},\boldsymbol{\xi}^{c,\text{DL}},\boldsymbol{\xi}^{c,\text{UL}},\boldsymbol{\xi}^s),\\
        \text{s.t.} \ \text{(\ref{eq:F})}&.
    \end{aligned}
\end{equation*}
For this subproblem, since \(\hat{\mathcal{G}}\) is a concave function w.r.t. $\mathbf{F}$, we can employ the Lagrange dual method to obtain the closed-form expression of \(\mathbf{F}\). The Lagrangian function is defined as 
\begin{equation}
\begin{aligned}
    \mathcal{L}(\mathbf{F}, \tau) &= -\hat{\mathcal{G}}(\mathbf{F}|\mathbf{p}_t, \mathbf{p}_r, \mathbf{w}_r, \mathbf{w}_s, \mathbf{f}_{\text{UL}}, \boldsymbol{\mu},\boldsymbol{\xi}^{c,\text{DL}},\boldsymbol{\xi}^{c,\text{UL}},\boldsymbol{\xi}^s) \\
    &+ \tau \left(\text{Tr}(\mathbf{F}^H\mathbf{F}) - P_{\text{DL}}\right).
\end{aligned}
\end{equation}
The corresponding Lagrangian dual problem is characterized by the Karush–Kuhn–Tucker (KKT) conditions as follows.
\begin{subequations}
    \begin{align}
        \frac{\partial \mathcal{L}(\mathbf{F}, \tau)}{\partial \mathbf{F}} &= 0, \label{eq:gradientF}\\
        \text{Tr} \left(\mathbf{F}^H\mathbf{F}\right) - P_0 &\leq 0, \\
        \tau &\geq 0, \\
        \tau \left(\text{Tr} \left(\mathbf{F}^H\mathbf{F}\right) - P_0\right) &= 0.
    \end{align}
\end{subequations}
These conditions yield a closed-form expression for the optimal solution of $\mathbf{F}$, where the $k$-th column is given by
\begin{equation}
\mathbf{f}_k(\tau) = \bigg(\big( \boldsymbol{\Lambda}_k^{T} + \tau \mathbf{I} \big)^{-1}\bigg)^{*} \boldsymbol{\varphi}_k,
\label{precoding_update}
\end{equation}
where $\boldsymbol\Lambda_k$ and $\boldsymbol\varphi_k$ are shown as (\ref{opt:f_lambda}) and (\ref{opt:f_phi}) respectively.

\begin{figure*}[t]
\small
\setlength\abovedisplayskip{3pt}
\setlength\belowdisplayskip{3pt}
\begin{align}
    &\boldsymbol{\Lambda}_k=\varpi_{c,\text{UL}} \sum^{K_{\text{UL}}}_{k=1} \Bigg\{  \sum_{c=1}^C \eta_c|\alpha_c|^2 \left[\mathbf{w}^H_{r,k}  \mathbf{b}_c(\mathbf{p}_r) \mathbf{a}_c^H(\mathbf{p}_t) \right]^H \left[\mathbf{w}^H_{r,k}  \mathbf{b}_c(\mathbf{p}_r) \mathbf{a}_c^H(\mathbf{p}_t) \right]  +   \eta_s|\alpha_s|^2 \left[\mathbf{w}^H_{r,k}  \mathbf{b}_s(\mathbf{p}_r) \mathbf{a}_s^H(\mathbf{p}_t) \right]^H \left[\mathbf{w}^H_{r,k}  \mathbf{b}_s(\mathbf{p}_r) \mathbf{a}_s^H(\mathbf{p}_t) \right]  \notag\\
    & +\left[\mathbf{w}_{r,k}^H \mathbf{H}_{\text{SI}}^H(\mathbf{p}_t,\mathbf{p}_r)\right]^H \left[\mathbf{w}^H_{r,k} \mathbf{H}_{\text{SI}}^H(\mathbf{p}_t,\mathbf{p}_r)\right]   \Bigg\}   +  \varpi_s \ ||\boldsymbol{\xi}^s||^2 \Bigg\{ \sum_{c=1}^C \eta_c|\alpha_c|^2 \left[\mathbf{w}^H_{s}  \mathbf{b}_c(\mathbf{p}_r) \mathbf{a}_c^H(\mathbf{p}_t) \right]^H \left[\mathbf{w}^H_{s}  \mathbf{b}_c(\mathbf{p}_r) \mathbf{a}_c^H(\mathbf{p}_t) \right] \notag\\
    &+   \eta_s|\alpha_s|^2 \left[\mathbf{w}^H_{s}  \mathbf{b}_s(\mathbf{p}_r) \mathbf{a}_s^H(\mathbf{p}_t) \right]^H \left[\mathbf{w}^H_{s}  \mathbf{b}_s(\mathbf{p}_r) \mathbf{a}_s^H(\mathbf{p}_t) \right]   +\left[\mathbf{w}_{s}^H \mathbf{H}_{\text{SI}}^H(\mathbf{p}_t,\mathbf{p}_r)\right]^H \left[\mathbf{w}^H_{s} \mathbf{H}_{\text{SI}}^H(\mathbf{p}_t,\mathbf{p}_r)\right]   \Bigg\}      +\varpi_{c,\text{DL}} |\xi_k^{c,\text{DL}}|^2  \mathbf{h}_k\mathbf{h}_k^H   \label{opt:f_lambda}, \\
&\boldsymbol{\varphi}_k=\varpi_{c,\text{DL}}\sqrt{1+\mu_k} {\xi_k^{c,\text{DL}}}^*\mathbf{h}_k(\mathbf{p}_t) +\varpi_s\sqrt{1+\mu_{K_{\text{DL}}+K_{\text{UL}}+1}}\sqrt{\eta_s}\alpha_s^{*}\xi^{s*}_k\mathbf{a}_s(\mathbf{p}_t)\mathbf{b}^H_s(\mathbf{p}_r)\mathbf{w}_s \label{opt:f_phi},\\
    &\bm{\gamma}_{r,k}= \sqrt{1+\mu_{K_{\text{DL}}+k}} {\xi^{c,\text{UL}}_k}^* f_{\text{UL},k}^* \mathbf{h}_{\text{UL},k}(\mathbf{p}_r)    \label{opt:w_gamma_r} ,\\
        &\bm{\Psi}_{r,k} = |\boldsymbol{\xi}^{c,\text{UL}}_k|^2   \bigg\{ \sum_{c=1}^{C}\eta_c|\alpha_c|^2\Big[ \mathbf{b}_c(\mathbf{p}_r)\mathbf{a}_c^H(\mathbf{p}_t)\mathbf{F}\Big]\Big[ \mathbf{b}_c(\mathbf{p}_r)\mathbf{a}_c^H(\mathbf{p}_t)\mathbf{F}\Big]^H + \eta_s|\alpha_s|^2\Big[ \mathbf{b}_s(\mathbf{p}_r)\mathbf{a}_s^H(\mathbf{p}_t)\mathbf{F}\Big]\Big[ \mathbf{b}_s(\mathbf{p}_r)\mathbf{a}_s^H(\mathbf{p}_t)\mathbf{F}\Big]^H   \notag \\ 
      & \quad\quad\quad+ \sum\limits_{j=1 }^{K_{\text{UL}}} |f_{\text{UL},j}|^2 \mathbf{h}_{\text{UL},j}^H \mathbf{h}_{\text{UL},j} +\Big[\mathbf{H}_{\text{SI}}^H(\mathbf{p}_t,\mathbf{p}_r)\mathbf{F}\Big]\Big[\mathbf{H}_{\text{SI}}^H(\mathbf{p}_t,\mathbf{p}_r)\mathbf{F}\Big]^H + \sigma_s^2 \mathbf{I} \bigg\}  \label{opt:w_psi_r} ,  \\
          &\bm{\gamma}_s= \sqrt{1+\mu_{K_{\text{DL}}+K_{\text{UL}}+1}} \sqrt{\eta_s}\alpha_s^*   {\boldsymbol{\xi}^s}^H \mathbf{F}^H \mathbf{a}_s(\mathbf{p}_t) \mathbf{b}_s^H(\mathbf{p}_r) \label{opt:w_gamma_s}, \\
    &\bm{\Psi}_s =  ||\boldsymbol{\xi}^s||^2   \bigg\{ \sum_{c=1}^{C}\eta_c|\alpha_c|^2\Big[ \mathbf{b}_c(\mathbf{p}_r)\mathbf{a}_c^H(\mathbf{p}_t)\mathbf{F}\Big]\Big[ \mathbf{b}_c(\mathbf{p}_r)\mathbf{a}_c^H(\mathbf{p}_t)\mathbf{F}\Big]^H +\eta_s|\alpha_s|^2\Big[ \mathbf{b}_s(\mathbf{p}_r)\mathbf{a}_s^H(\mathbf{p}_t)\mathbf{F}\Big]\Big[ \mathbf{b}_s(\mathbf{p}_r)\mathbf{a}_s^H(\mathbf{p}_t)\mathbf{F}\Big]^H \notag \\
    & \quad\quad\quad+ \sum\limits_{j=1 }^{K_{\text{UL}}} |f_{\text{UL},j}|^2 \mathbf{h}_{\text{UL},j}^H \mathbf{h}_{\text{UL},j} + \Big[\mathbf{H}_{\text{SI}}^H(\mathbf{p}_t,\mathbf{p}_r)\mathbf{F}\Big]\Big[\mathbf{H}_{\text{SI}}^H(\mathbf{p}_t,\mathbf{p}_r)\mathbf{F}\Big]^H + \sigma_s^2 \mathbf{I} \bigg\}  \label{opt:w_psi_s} .    
\end{align}
\hrulefill
\end{figure*}

\begin{algorithm}[t]
    \renewcommand{\algorithmicrequire}{\textbf{Initialization:}}
	\renewcommand{\algorithmicensure}{\textbf{Output:}}
    \caption{Iterative optimization for transmit and receive beamforming matrices.}
    \label{param_all}
    \begin{algorithmic}[1]
        \REQUIRE Choose the upper bound and lower bound of $\tau$ and $\tau_u$ as $\tau_{\text{max}}$,$\tau_{\text{min}}$,$\tau_{u,\text{max}}$ and $\tau_{u,\text{min}}$, tolerence $\epsilon$, power limit of downlink transmit $P_{\text{DL}}$ and power limit of uplink transmit $P_{\text{UL}}$; randomly initial $\boldsymbol{\xi}^{c,\text{DL}}$,$\boldsymbol{\xi}^{c,\text{UL}}$,$\boldsymbol{\xi}^s$, $\boldsymbol{\mu}$, $\mathbf{w}_r$, $\mathbf{w}_s$, set iteration index $i=1$.
        \REPEAT
            \REPEAT \label{bisection_st}
                \STATE Compute $\tau=(\tau_{\text{max}}+\tau_{\text{min}})/2$. 
                \STATE Update $\mathbf{F}^{(i)}$ as (\ref{precoding_update}).  
                \STATE Compute power $P$ of $\mathbf{F}^{(i)}$. 
                \STATE \textbf{if} { $P > P_{\text{DL}}$ } \textbf{then} $\tau_{\text{min}}=\tau$ \textbf{else} $\tau_{\text{max}}=\tau$.
            \UNTIL{$\left| P - P_{\text{DL}} \right|< \epsilon$.}

            \REPEAT
                \STATE Compute $\tau_u=(\tau_{u,\text{max}}+\tau_{u,\text{min}})/2$. 
                \STATE Update $\mathbf{f}_{\text{UL}}^{(i)}$ as (\ref{fUL_update}).
                \STATE Compute power $P$ of $\mathbf{f}_{\text{UL}}^{(i)}$. 
                \STATE \textbf{if} { $P > P_{\text{UL}}$ } \textbf{then} $\tau_{u,\text{min}}=\tau_u$ \textbf{else} $\tau_{u,\text{max}}=\tau_u$.
            \UNTIL{$\left|P - P_{\text{UL}} \right|< \epsilon$.}  \label{bisection_end}
            \STATE Update $\mathbf{w}_r^{(i)},\mathbf{w}_s^{(i)},\boldsymbol{\mu}^{(i)}, 
            {\boldsymbol{\xi}^{c,\text{DL}}}^{(i)},{\boldsymbol{\xi}^{c,\text{UL}}}^{(i)},{\boldsymbol{\xi}^{s}}^{(i)}$  as(\ref{wr_update}), (\ref{ws_update}), (\ref{mu_update}),
            (\ref{ksic_DL_update}), (\ref{ksic_UL_update}), (\ref{ksis_update}), separately. Set iteration index $i=i+1$.
        \UNTIL{the value of objective function converge.}
        \ENSURE $\mathbf{F}^{(i-1)},\mathbf{w}^{(i-1)}_s,\mathbf{w}^{(i-1)}_r,\mathbf{f}_{\text{UL}}^{(i-1)}$.
    \end{algorithmic}
\end{algorithm}

Similarly, the subproblem for optimizing the uplink user precoding coefficients can be formulated as
\begin{equation*}
    \begin{aligned}
        (\text{SP.2}) \max_{\mathbf{f}_{\text{UL}}}\hat{\mathcal{G}}(\mathbf{f}_{\text{UL}}|\mathbf{F}, \mathbf{p}_t, \mathbf{p}_r, \mathbf{w}_r, \mathbf{w}_s,\boldsymbol{\mu},\boldsymbol{\xi}^{c,\text{DL}},\boldsymbol{\xi}^{c,\text{UL}},\boldsymbol{\xi}^s),&~ \\
         \text{s.t.}  \text{(\ref{eq:UL_P})}&.
    \end{aligned}
\end{equation*}
We also construct the corresponding Lagrangian function with respect to $\mathbf{f}_{\text{UL}}$ as 
\begin{equation}
\begin{aligned}
        \mathcal{L}(\mathbf{f}_{\text{UL}}, \tau_u) &= -\hat{\mathcal{G}}(\mathbf{f}_{\text{UL}}|\mathbf{F}, \mathbf{p}_t, \mathbf{p}_r, \mathbf{w}_r, \mathbf{w}_s, \boldsymbol{\mu},\boldsymbol{\xi}^{c,\text{DL}},\boldsymbol{\xi}^{c,\text{UL}},\boldsymbol{\xi}^s) \\
    &+ \tau_u \left(\text{Tr}(\mathbf{f}_{\text{UL}}^H \mathbf{f}_{\text{UL}} ) - P_{\text{UL}}\right).
\end{aligned}
\end{equation}
Based on the KKT conditions, the closed-form solution for the $k$-th element of $\mathbf{f}_{\text{UL}}$ is given by
\begin{equation}
    f_{\text{UL},k}(\tau_u) = \big(\frac{\varphi_{\text{UL},k}}{\Lambda_{\text{UL},k} + \tau_u  } \big)^*,
    \label{fUL_update}
\end{equation}
where
\begin{equation}
\small
\begin{aligned}
    \Lambda_{\text{UL},k} \!= \! \varpi_s ||\boldsymbol{\xi}^s||^2 |\mathbf{w}_{s}^H \mathbf{h}_{\text{UL},k}^H|^2 + \varpi_{c,\text{UL}}  \sum_{j=1}^{K_{\text{UL}}} |\xi^{c,\text{UL}}_j \mathbf{w}_{r,j}^H \mathbf{h}_{\text{UL},k}^H|^2,
\end{aligned}
\end{equation}
\begin{equation}
    \varphi_{\text{UL},k}=\varpi_{c,\text{UL}} \sqrt{1+\mu_{K_{\text{DL}}+k}} \xi^{c,\text{UL}}_k \mathbf{w}^H_{r,k} \mathbf{h}_{\text{UL},k}^H(\mathbf{p}_r).
\end{equation}
$\tau$ and $\tau_u$ should be chosen to satisfy the dual feasibility condition and the complementary slackness condition. Here we adopt a bisection method \cite{bisection_ref} to search for the appropriate solution. The detailed procedure is described in Step \ref{bisection_st} to \ref{bisection_end} of \textbf{Algorithm \ref{param_all}}.

For receive beamforming matrices $\mathbf{w}_r$ and $\mathbf{w}_s$, we have
\begin{equation*}
    \begin{aligned}
         &(\text{SP.3}) \max_{\mathbf{w}_r}\hat{\mathcal{G}}(\mathbf{w}_r|\mathbf{F}, \mathbf{p}_t, \mathbf{p}_r, \mathbf{w}_s,\mathbf{f}_{\text{UL}}, \boldsymbol{\mu},\boldsymbol{\xi}^{c,\text{DL}},\boldsymbol{\xi}^{c,\text{UL}},\boldsymbol{\xi}^s),\\
        &(\text{SP.4}) \max_{\mathbf{w}_s}\hat{\mathcal{G}}(\mathbf{w}_s|\mathbf{F}, \mathbf{p}_t, \mathbf{p}_r, \mathbf{w}_r,\mathbf{f}_{\text{UL}}, \boldsymbol{\mu},\boldsymbol{\xi}^{c,\text{DL}},\boldsymbol{\xi}^{c,\text{UL}},\boldsymbol{\xi}^s).
    \end{aligned}
\end{equation*}
Since they have no constraint and \(\hat{\mathcal{G}}\) is a concave function w.r.t. $\mathbf{w}_r$ and $\mathbf{w}_s$, we could directly derive the closed form of $\mathbf{w}_r$ and $\mathbf{w}_s$ as 
\begin{equation}
    \mathbf{w}_{r,k}=\big(\bm{\gamma}_{r,k} \bm{\Psi}_{r,k}^{-1}\big)^H,
    \label{wr_update}
\end{equation}
\begin{equation}
    \mathbf{w}_{s}=\big(\bm{\gamma}_{s} \bm{\Psi}_{s}^{-1}\big)^H,
    \label{ws_update}
\end{equation}
where $\bm{\gamma}_{r,k},\bm{\Psi}_{r,k},\bm{\gamma}_{s},\bm{\Psi}_{s}$ are shown as (\ref{opt:w_gamma_r}), (\ref{opt:w_psi_r}), (\ref{opt:w_gamma_s}) and (\ref{opt:w_psi_s}) respectively.

\subsection{Auxiliary Variables Optimization}
To optimize the auxiliary variables, we consider the following subproblems
\begin{equation*}
    \begin{aligned}
    &(\text{SP.5}) \max_{\boldsymbol{\mu}}\hat{\mathcal{G}}(\boldsymbol{\mu}|\mathbf{F}, \mathbf{p}_t, \mathbf{p}_r, \mathbf{w}_r, \mathbf{w}_s, \mathbf{f}_{\text{UL}}, \boldsymbol{\xi}^{c,\text{DL}},\boldsymbol{\xi}^{c,\text{UL}},\boldsymbol{\xi}^s),\\
        &(\text{SP.6}) \max_{\boldsymbol{\xi}^{c,\text{DL}}\!,\boldsymbol{\xi}^{c,\text{UL}}\!,\boldsymbol{\xi}^s}\hat{\mathcal{G}}(\boldsymbol{\xi}^{c,\text{DL}},\!\boldsymbol{\xi}^{c,\text{UL}},\!\boldsymbol{\xi}^s\!|\mathbf{F}, \mathbf{p}_t, \mathbf{p}_r, \mathbf{w}_r, \mathbf{w}_s, \mathbf{f}_{\text{UL}}, \boldsymbol{\mu}) \label{eq:xisxic}.
    \end{aligned}
\end{equation*}
We need to update $\boldsymbol{\mu}$  with $ \frac{\partial{\mathcal{G}}_l}{\partial \boldsymbol{\mu}} = 0$, then we can derive the optimal $\mathcal{G}_l$ as  $\mathcal{G}$ since they are Lagrangian dual.
Therefore, $\boldsymbol{\mu}$ is updated with different $k$ as
\begin{equation}
    \footnotesize
    \mu_k \! = \!
    \left\{  \!
    \begin{aligned}
        & \frac{|\mathbf{h}_{\text{DL},k}^H(x)\mathbf{f}_k|^2}{\sum\limits_{j=1, j \neq k}^{K_{\text{DL}}} |\mathbf{h}_{\text{DL},k}^H(x)\mathbf{f}_j|^2 + \sum\limits_{j=1}^{K_{\text{UL}}} |g_{j,k}|^2+\sigma_k^2},  k \in \{1, \dots, K_{\text{DL}}\},\\
        &\frac{| \mathbf{w}^H_{r,k} \mathbf{h}_{\text{UL},k}^H(\mathbf{p}_r) f_{\text{UL},k} |^2 } {S_{r,k}\!+\!C_{r,k}\!+\!SI_{r,k} \! + \! \sum\limits_{j=1,j\neq k }^{K_{\text{UL}}} \! | \mathbf{w}^H_{r,k} \mathbf{h}_{\text{UL},j}^H(\mathbf{p}_r) f_{\text{UL},j} |^2 \!+ \! ||\mathbf{w}_{r,k}||^2\sigma_s^2}\\
        &\quad\quad\quad\quad\quad\quad\quad\quad\quad\quad\quad ,k \in \{K_{\text{DL}}+1, \dots, K_{\text{DL}}+K_{\text{UL}} \},\\
        &\frac{S_s}{C_s + SI_s + \sum\limits_{j=1}^{K_{\text{UL}}} | \mathbf{w}^H_s \mathbf{h}_{\text{UL},j}^H(\mathbf{p}_r) f_{\text{UL},j} |^2 + ||\mathbf{w}_s||^2 \sigma_s^2}\\
    &\quad\quad\quad\quad\quad\quad\quad\quad\quad\quad\quad ,k=K_{\text{DL}}+K_{\text{UL}}+1.\label{mu_update}
    \end{aligned}
    \right.
\end{equation}

Since \(\hat{\mathcal{G}}\) is concave w.r.t $\boldsymbol{\xi^{c,\text{DL}}}$, $\boldsymbol{\xi^{c,\text{UL}}}$ and $\boldsymbol{\xi}^s$. We could obtain the closed-form solutions by setting the partial derivatives to zero, i.e., 
 $ \frac{\partial\hat{\mathcal{G}}(\boldsymbol{\xi}^{c,\text{DL}},\boldsymbol{\xi}^{c,\text{UL}},\boldsymbol{\xi}^s|\mathbf{F}, \mathbf{p}_t, \mathbf{p}_r, \mathbf{w}_r, \mathbf{w}_s, \mathbf{f}_{\text{UL}}, \boldsymbol{\mu})}{\partial \boldsymbol{\xi}^{c,\text{DL}}} = 0$, $ \frac{\partial\hat{\mathcal{G}}(\boldsymbol{\xi}^{c,\text{DL}},\boldsymbol{\xi}^{c,\text{UL}},\boldsymbol{\xi}^s|\mathbf{F}, \mathbf{p}_t, \mathbf{p}_r, \mathbf{w}_r, \mathbf{w}_s, \mathbf{f}_{\text{UL}}, \boldsymbol{\mu})}{\partial \boldsymbol{\xi}^{c,\text{UL}}} = 0$ and $~~~~~~~~~~$
 $ \frac{\partial\hat{\mathcal{G}}(\boldsymbol{\xi}^{c,\text{DL}},\boldsymbol{\xi}^{c,\text{UL}},\boldsymbol{\xi}^s|\mathbf{F}, \mathbf{p}_t, \mathbf{p}_r, \mathbf{w}_r, \mathbf{w}_s, \mathbf{f}_{\text{UL}}, \boldsymbol{\mu})}{\partial \boldsymbol{\xi}^s} = 0$.
Therefore, the closed form of $\xi^{c,\text{DL}}_k$ is 
\begin{equation}
    \xi^{c,\text{DL}}_k = \frac{\sqrt{1+\mu_{k}}\mathbf{f}_k^H \mathbf{h}_{\text{DL},k} \left(\mathbf{p}_t\right)}{\sum_{j=1}^{K_{\text{DL}}} |\mathbf{h}_{\text{DL},k}^H(\mathbf{p}_t) \mathbf{f}_j|^2 + \sum\nolimits_{j=1}^{K_{\text{UL}}} |g_{j,k}|^2 +\sigma_k^2}.
    \label{ksic_DL_update}
\end{equation}
Similarly, the closed-form solution for $\xi^{c,\text{UL}}_k$ can be derived as
\begin{equation}
\scalebox{0.85}{$
    \begin{aligned}
    \xi^{c,\text{UL}}_k = \frac{\sqrt{1+\mu_{K_{\text{DL}}+k}} f_{\text{UL},j}^H \mathbf{h}_{\text{UL},j}(\mathbf{p}_r) \mathbf{w}_{r,k}}{C_{r,k} + S_{r,k} +SI_{r,k} + \sum\limits_{j=1}^{K_{\text{UL}}} |f_{\text{UL},j} \mathbf{w}_{r,k}^H \mathbf{h}_{\text{UL},j}^H(\mathbf{p}_r)  |^2 + ||\mathbf{w}_{r,k}||^2 \sigma_s^2},
    \label{ksic_UL_update}
    \end{aligned}
    $}
\end{equation}
and for $\boldsymbol{\xi}^s$ we have
\begin{equation}
        \small
        \boldsymbol{\xi}^{s} \! = \! \frac{\sqrt{1+\mu_{K_{\text{DL}}+K_{\text{UL}}+1}}\left(\sqrt{\eta_s}\alpha_s \mathbf{w}^H_s \mathbf{b}_s(\mathbf{p}_r)\mathbf{a}_s^H(\mathbf{p}_t)\mathbf{F}\right)^H}{C_s + S_s +SI_s + \sum\limits_{j=1}^{K_{\text{UL}}} | f_{\text{UL},j} \mathbf{w}^H_s \mathbf{h}_{\text{UL},j}^H(\mathbf{p}_r)  |^2 +||\mathbf{w}_{s}||^2\sigma_s^2}.
        \label{ksis_update}
\end{equation}

\textbf{Algorithm~\ref{param_all}} summarizes the iterative optimization procedure for both the beamforming matrices and the auxiliary variables. By executing \textbf{Algorithm~\ref{param_all}}, a locally optimal solution can be obtained for the given antenna positions.

\subsection{Antenna Position Optimization}
In the previous subsections, we derive the optimization
algorithm with fixed antenna positions. 
To further enhance the performance of the system by leveraging movable antenna, we need to address the following subproblems with given beamforming matrices.
\begin{equation}
    \begin{aligned}
        (\text{SP.7}) \max_{\mathbf{p}_t}\hat{\mathcal{G}}(\mathbf{p}_t|\mathbf{F}, \mathbf{p}_r, \mathbf{w}_r, \mathbf{w}_s, \mathbf{f}_{\text{UL}}, \boldsymbol{\mu},\boldsymbol{\xi}^{c,\text{DL}},\boldsymbol{\xi}^{c,\text{UL}},\boldsymbol{\xi}^s),& \\ 
        \text{s.t.} \ \text{(\ref{t_range_constraint}),\ (\ref{distance_constraint})} &. \notag
    \end{aligned}
\end{equation}
\begin{equation}
    \begin{aligned}
        (\text{SP.8}) \max_{\mathbf{p}_r}\hat{\mathcal{G}}(\mathbf{p}_r|\mathbf{F}, \mathbf{p}_t,  \mathbf{w}_r, \mathbf{w}_s, \mathbf{f}_{\text{UL}}, \boldsymbol{\mu},\boldsymbol{\xi}^{c,\text{DL}},\boldsymbol{\xi}^{c,\text{UL}},\boldsymbol{\xi}^s),&\\
        \text{s.t.} \ \text{(\ref{r_range_constraint}),\  (\ref{distance_constraint})} &.\notag
    \end{aligned}
\end{equation}
Due to the non-convex nature of these subproblems, deriving global optimal solutions is generally intractable. Therefore, we adopt a GA approach to iteratively search for local optimal solutions for both the transmit and receive antenna positions.

In the $i$-th iteration of the GA process, the $m$-th transmit antenna's position is updated as 
\begin{equation}
    \footnotesize
    \mathbf{p}_{t,m}^{(i)} \!= \! \mathbf{p}_{t,m}^{(i-1)} \! + \! \boldsymbol{\delta}_{t} \nabla_{\mathbf{p}_{t,m}}\hat{\mathcal{G}}(\mathbf{p}_t|\mathbf{F}, \mathbf{p}_r, \mathbf{w}_r, \mathbf{w}_s, \mathbf{f}_{\text{UL}}, \boldsymbol{\mu},\boldsymbol{\xi}^{c,\text{DL}},\boldsymbol{\xi}^{c,\text{UL}},\boldsymbol{\xi}^s) , \label{transmit_antenna_GA}
\end{equation}
where $\boldsymbol{\delta}_t$ contains the step size for GA along the x-axis and y-axis. For the gradient along the x-axis, it is shown as (\ref{opt:gra_xt}).
\begin{figure*}[t]
\small
\setlength\abovedisplayskip{3pt}
\setlength\belowdisplayskip{3pt}
\begin{align}
&\nabla_{x_t}\hat{\mathcal{G}}(\mathbf{F}, \mathbf{p}_t, \mathbf{p}_r, \mathbf{w}_r, \mathbf{w}_s ,\mathbf{f}_{\text{UL}}, 
\boldsymbol{\mu},\boldsymbol{\xi}^{c,\text{DL}},\boldsymbol{\xi}^{c,\text{UL}},\boldsymbol{\xi}^s)= \varpi_{c,\text{DL}}\sum_{k=1}^{K_{\text{DL}}} \bigg[   2\sqrt{1+\mu_k} \text{Re}( \xi_k^{c,\text{DL}} \mathbf{f}_k^H       \nabla_{x_t}\mathbf{h}_{\text{DL},k}) - |\xi_s^{c,\text{DL}}|^2 \sum_{j=1}^{K_{\text{DL}}} 2 \text{Re}(\mathbf{h}_{\text{DL},k}^H\mathbf{f}_j \mathbf{f}_j^H  \nabla_{x_t}\mathbf{h}_{\text{DL},k}  )   \bigg]  \notag\\   
  &-2\varpi_{c,\text{UL}} \sum_{k=1}^{K_{\text{UL}}} |\xi^{c,\text{UL}}_k|^2  \bigg[ \sum_{c=1}^{C} \eta_c|\alpha_c|^2 \text{Re} ( \mathbf{b}_c^H \mathbf{w}_{r,k} \mathbf{w}^H_{r,k} \mathbf{b}_c^H \mathbf{a}_c^H \mathbf{F} \mathbf{F}^H 
 \nabla_{x_t}\mathbf{a}_c )    +  \eta_s|\alpha_s|^2\text{Re} ( \mathbf{b}_s^H \mathbf{w}_{r,k} \mathbf{w}^H_{r,k} \mathbf{b}_s^H \mathbf{a}_s^H \mathbf{F} \mathbf{F}^H  
  \nabla_{x_t}\boldsymbol{\alpha}_s )    \notag \\
  & + \text{Re} ( \mathbf{w}_{r,k} \mathbf{w}_{r,k}^H  \mathbf{H}_{\text{SI}}^H \mathbf{F}  \mathbf{F}^H  \nabla_{x_t}\mathbf{H}_{\text{SI}}  )     \bigg]    + 2 \varpi_s \bigg\{ \sqrt{1+\mu_{K_{\text{DL}}+K_{\text{UL}}+1}} \text{Re} ( \sqrt{\eta_k} \alpha_k \boldsymbol{\xi}^s \mathbf{F} \mathbf{w}_s^H \mathbf{b}_k \nabla_x\mathbf{a}_s )   \notag\\
    &  - ||\boldsymbol{\xi}^s||^2 \bigg[  \sum_{c=1}^{C} \eta_c|\alpha_c|^2 \text{Re} ( \mathbf{b}_c^H \mathbf{w}_{s} \mathbf{w}^H_{s} \mathbf{b}_c^H \mathbf{a}_c^H \mathbf{F} \mathbf{F}^H 
 \nabla_{x_t}\mathbf{a}_c )  +  \eta_s|\alpha_s|^2\text{Re} ( \mathbf{b}_s^H \mathbf{w}_{s} \mathbf{w}^H_{s} \mathbf{b}_s^H \mathbf{a}_s^H \mathbf{F} \mathbf{F}^H  
  \nabla_{x_t}\mathbf{a}_s ) + \text{Re}( \mathbf{w}_s \mathbf{w}_s^H  \mathbf{H}_{\text{SI}}^H \mathbf{F}  \mathbf{F}^H  \nabla_{x_t}\mathbf{H}_{\text{SI}} ) \bigg]   \bigg\}. \label{opt:gra_xt}\\
    &\nabla_{x_r}\hat{\mathcal{G}}(\mathbf{F}, \mathbf{p}_t, \mathbf{p}_r, \mathbf{w}_r, \mathbf{w}_s ,\mathbf{f}_{\text{UL}}, \boldsymbol{\mu},\boldsymbol{\xi}^{c,\text{DL}},\boldsymbol{\xi}^{c,\text{UL}},\boldsymbol{\xi}^s)= \varpi_{c,\text{UL}} \sum_{k=1}^{K_{\text{UL}}}\bigg\{ 2\sqrt{1+\mu_{K_{\text{DL}}+k}} \text{Re} ( f_{\text{UL},k} \boldsymbol{\xi}_k^{c,\text{UL}} \mathbf{w}_{r,k}^H \nabla_{x_r}\mathbf{h}_{\text{UL},k} )  \notag\\ 
    & - |\boldsymbol{\xi}_k^{c,\text{UL}}|^2 \bigg[  \sum_{c=1}^{C} \eta_c |\alpha_c|^2 \text{Re}( \mathbf{a}_c^H \mathbf{F} \mathbf{F}^H  \mathbf{a}_c \mathbf{b}_c^H \mathbf{w}_{r,k} \mathbf{w}^H_{r,k} \nabla_{x_r}\mathbf{b}_c    + \eta_s |\alpha_s|^2 \text{Re}( \mathbf{a}_s^H \mathbf{F} \mathbf{F}^H  \mathbf{a}_s \mathbf{b}_s^H \mathbf{w}_{r,k} \mathbf{w}^H_{r,k} \nabla_{x_r}\mathbf{b}_s)  + \text{Re}( \mathbf{w}_{r,k} \mathbf{w}_{r,k}^H  \mathbf{H}_{\text{SI}}^H \mathbf{F}  \mathbf{F}^H  \nabla_{x_r}\mathbf{H}_{\text{SI}} )  \notag\\
&+ \sum_{j=1}^{K_{\text{UL}}} |f_{\text{UL},j}|^2 \text{Re} ( \mathbf{w}_{r,k} \mathbf{w}_{r,k}^H \mathbf{h}^H_{\text{UL},j} \nabla_{x_r} \mathbf{h}_{\text{UL},j} )
    \bigg]  \bigg\}   +\varpi_s \bigg\{ \sqrt{1+\mu_{K_{\text{DL}}+K_{\text{UL}}+1}} \text{Re} ( \sqrt{\eta_s} \alpha_s \mathbf{a}_s^H
\mathbf{F} \boldsymbol{\xi}^s \mathbf{w}_s^H \nabla_{x_r}\mathbf{b}_s
    )       \notag\\
& -   ||\boldsymbol{\xi}^s||^2 \bigg[ \sum_{c=1}^{C} \eta_c |\alpha_c|^2 \text{Re}( \mathbf{a}_c^H \mathbf{F} \mathbf{F}^H  \mathbf{a}_c \mathbf{b}_c^H \mathbf{w}_{r,k} \mathbf{w}^H_{r,k} \nabla_{x_r}\mathbf{b}_c ) + \eta_s |\alpha_s|^2 \text{Re}( \mathbf{a}_s^H \mathbf{F} \mathbf{F}^H  \mathbf{a}_s \mathbf{b}_s^H \mathbf{w}_{r,k} \mathbf{w}^H_{r,k} \nabla_{x_r}\mathbf{b}_s )  +  \text{Re}( \mathbf{w}_s \mathbf{w}_s^H  \mathbf{H}_{\text{SI}}^H \mathbf{F}  \mathbf{F}^H  \nabla_{x_r}\mathbf{H}_{\text{SI}} )  \notag \\ 
&+ \sum_{j=1}^{K_{\text{UL}}} |f_{\text{UL},j}|^2 \text{Re} (\mathbf{w}_s \mathbf{w}_s^H \mathbf{h}^H_{\text{UL},j} \nabla_{x_r} \mathbf{h}_{\text{UL},j} )  \bigg]  \bigg\} .   \label{opt:gra_xr}
\end{align}
\hrulefill
\end{figure*}

We could also derive the gradient along the y-axis by replacing $\nabla_{x_t}\mathbf{h}_{\text{DL},k}$, $\nabla_{x_t}\mathbf{a}_{c}$, $\nabla_{x_t}\mathbf{a}_{s}$, and $\nabla_{x_t}\mathbf{H}_{SI}$ in (\ref{opt:gra_xt}) with $\nabla_{y_t}\mathbf{h}_{\text{DL},k}$, $\nabla_{y_t}\mathbf{a}_{c}$, $\nabla_{y_t}\mathbf{a}_{s}$, and $ \nabla_{y_t}\mathbf{H}_{SI}$ respectively. 

For the $n$-th receive antenna, the position is updated as 
\begin{equation}
    \footnotesize
    \mathbf{p}_{r,n}^{(i)} \! = \! \mathbf{p}_{r,n}^{(i-1)} \!+\! \boldsymbol{\delta}_{r} \nabla_{\mathbf{p}_{r,n}}\hat{\mathcal{G}}(\mathbf{p}_r|\mathbf{F}, \mathbf{p}_t, \mathbf{w}_r, \mathbf{w}_s, \mathbf{f}_{\text{UL}}, \boldsymbol{\mu},\boldsymbol{\xi}^{c,\text{DL}},\boldsymbol{\xi}^{c,\text{UL}},\boldsymbol{\xi}^s) , \label{receive_antenna_GA}
\end{equation}
where $\boldsymbol{\delta}_{r}$ is the step size for GA along the x-axis and y-axis. For the gradient along the x-axis, it is shown as (\ref{opt:gra_xr}). 
 And for the gradient along the y-axis, we could derive by replacing 
 $\nabla_{x_r}\mathbf{h}_{\text{UL},k}$, $\nabla_{x_r}\mathbf{b}_{c}$, $\nabla_{x_r}\mathbf{b}_{s}$, and $\nabla_{x_r}\mathbf{H}_{SI}$ in (\ref{opt:gra_xr})
 with $\nabla_{y_r}\mathbf{h}_{\text{UL},k}$, $\nabla_{y_r}\mathbf{b}_{c}$, $\nabla_{y_r}\mathbf{b}_{s}$, and $\nabla_{y_r}\mathbf{H}_{SI}$ respectively.

To satisfy the constraints (\ref{t_range_constraint}), (\ref{r_range_constraint}), and (\ref{distance_constraint}), we need to verify compliance after repositioning the antennas.
If the constraints are not fulfilled, we would multiply the step size by 0.9 and regenerate the antenna positions until the constraints are satisfied. This could guarantee that the constraints are met during the GA process.

The AO-based algorithm is summarized as \textbf{Algorithm \ref{AObased_algorithm}}. We alternatively update the position of transmit and receive antennas and then derive the beamforming matrices with the new antenna positions using \textbf{Algorithm \ref{param_all}}. By leveraging the GA method to adjust the antenna positions, we capitalize on the flexibility of the movable antenna to enhance the system performance.

\begin{algorithm}[t] 
    \renewcommand{\algorithmicrequire}{\textbf{Initialization:}}
	\renewcommand{\algorithmicensure}{\textbf{Output:}}
    \caption{AO-based algorithm for optimization.}
    \label{AObased_algorithm}
    \begin{algorithmic}[1]
        \REQUIRE Randomly generate $\mathbf{F}^{(1)},\mathbf{w}_s^{(1)},\mathbf{w}_r^{(1)}$ and $\mathbf{f}_{\text{UL}}^{(1)}$, set iteration index $j=1$.
        \STATE Update $\mathbf{F}^{(1)},\mathbf{w}_s^{(1)},\mathbf{w}_r^{(1)}$ and $\mathbf{f}_{\text{UL}}^{(1)}$ as \textbf{Algorithm \ref{param_all}}.
        \REPEAT
            \REPEAT
            \STATE Update $\mathbf{p}_t^{(j)}$ as (\ref{transmit_antenna_GA})
            \WHILE{$\mathbf{p}_t^{(j)}$ do not satisfy constraint (\ref{t_range_constraint}) and (\ref{distance_constraint})}
                \STATE Adjust step size by $\boldsymbol{\delta}_t=0.9 \boldsymbol{\delta}_t$.
                \STATE Update $\mathbf{p}_t^{(j)}$ as (\ref{transmit_antenna_GA}). 
            \ENDWHILE
            \UNTIL{the value of objective function converge.}
            \REPEAT
            \STATE Update $\mathbf{p}_r^{(j)}$ as (\ref{receive_antenna_GA}).
            \WHILE{$\mathbf{p}_r^{(j)}$ do not satisfy constraint (\ref{r_range_constraint}) and (\ref{distance_constraint})}
                \STATE Adjust step size by $\boldsymbol{\delta}_r=0.9 \boldsymbol{\delta}_r$. 
                \STATE Update $\mathbf{p}_r^{(j)}$ as (\ref{receive_antenna_GA}). 
            \ENDWHILE
            \UNTIL{the value of objective function converge.}

            \STATE Update $\mathbf{F}^{(j)},\mathbf{w}_s^{(j)},\mathbf{w}_r^{(j)}$ and $\mathbf{f}_{\text{UL}}^{(j)}$ as \textbf{Algorithm \ref{param_all}}.
            \STATE Set $j=j+1$.
            
        \UNTIL{the value of objective function converge.}
        \ENSURE $\mathbf{F}^{(j-1)}, \mathbf{w}_r^{(j-1)}, \mathbf{w}_r^{(j-1)}, \mathbf{f}_{\text{UL}}^{(j-1)}, \mathbf{p}_t^{(j-1)}, \mathbf{p}_t^{(j-1)}$.
        
    \end{algorithmic}
    
\end{algorithm}

\subsection{PSO-based algorithm for updating antenna position}

Although antenna positions can be optimized via GA, the performance of the AO-based algorithm remains highly sensitive to the initial placement of the antennas, which significantly limits its ability to escape local optimum. Simulation results reveal that antenna movements tend to be restricted within a narrow region, leaving much of the feasible space underutilized and resulting in a local optimal result. 
To tackle this problem, we propose a PSO-based algorithm that utilizes PSO as an efficient method to explore sub-optimal result across the entire feasible region.
Given that conventional PSO does not adequately explore optimal results in specific regions. We integrate GA within the algorithm to assist PSO in more effectively searching for local optima.
This integration enables the particles to move more efficiently, ultimately yielding a superior sub-optimal result, which is further discussed in the next section.

In the PSO-based algorithm, there are $N_p$ particles, each particle $ \tilde{\mathbf{p}}_n^{(i)}$ contains the position of all $N_t+N_r$ antennas which is shown as 
\begin{equation}
    \tilde{\mathbf{p}}_n^{(i)}=[\mathbf{p}^{(i)}_{t,1},\cdots,\mathbf{p}^{(i)}_{t,N_t},\mathbf{p}^{(i)}_{r,1},\cdots,\mathbf{p}^{(i)}_{r,N_r}].
\end{equation}
During the optimization process, each particle updates its position in every iteration based on a velocity vector. This velocity is affected by the local optimal result for that particle and the global optimal result for all particles, as well as the velocity of last movement. 
For the $n$-th  particle in $i$-th iteration, velocity $\mathbf{v}^{(i)}_n$ is defined as 
\begin{equation}
    \mathbf{v}^{(i)}_n=\omega \mathbf{v}^{(i-1)}_n + c_1 \tau_1 (\tilde{\mathbf{p}}_g-\tilde{\mathbf{p}}_n^{(i-1)} ) + c_2 \tau_2 (\tilde{\mathbf{p}}_{l,n}-\tilde{\mathbf{p}}_n^{(i-1)} ) ,
    \label{velocity_update}
\end{equation}
where $\tilde{\mathbf{p}}_g$ represents the best antenna positions among all $N_p$ particles and $\tilde{\mathbf{p}}_{l,n}$ represents the best antenna positions for the $n$-th particle. $c_1$ and $c_2$ are learning factors for global optimal and local optimal respectively, while $\tau_1$ and $\tau_2$ are random factors which follow $\tau_1,\tau_2 \sim \mathcal{U}(0,1)$. $\omega$ is the inertia weight, which could make the particle jump out of sub-optimal result at the initial stage and converge at the end stage. It is dynamically adjusted as a function of the iteration index and can be derived as
\begin{equation}
    \omega=\bigg(\omega_{\text{max}}-\frac{(\omega_{\text{max}}-\omega_{\text{min}})i}{I_p} \bigg),
\end{equation}
where $\omega_{\text{max}}$ and $\omega_{\text{min}}$ denote the initial and final values of the inertia weight, and $I_p$ is the total number of PSO iterations.
Subsequently, the antenna positions of each particle is updated according to the new velocity as
\begin{equation}
    \tilde{\mathbf{p}}_n^{(i)}=\tilde{\mathbf{p}}_n^{(i-1)}+\mathbf{v}^{(i)}_n . \label{positon_update}
\end{equation}

To obtain the best antenna positions among all particles, we need to evaluate the fitness of each particle. Here, we set the weighted sum of communication rate and sensing MI as the fitness of particles, which is
\begin{equation}
    R_n^{(i)}=\mathcal{G}(\mathbf{F}, \tilde{\mathbf{p}}_n^{(i)}, \mathbf{w}_r,\mathbf{w}_s,\mathbf{f}_{\text{UL}}).
\end{equation}
Let $R_g$ denotes the maximum fitness for all particles and $R_{l,n}$ denotes the maximum fitness for the $n$-th particle.

In order to satisfy constraint (\ref{t_range_constraint}) and (\ref{r_range_constraint}), we project the antennas that move out of the feasible region back to the bound along the x-axis and y-axis respectively with bounding function $B\{\cdot\}$ as 
\begin{equation}
    \quad[B\{\tilde{\mathbf{p}}_n^{(i)}\}]_{1,k}=
    \left\{  
    \begin{aligned}
         & X_{\text{MIN}} , \text{if} \; [\tilde{\mathbf{p}}_n^{(i)}]_{1,k} < X_{\text{MIN}},\\
         & X_{\text{MAX}} , \text{if} \; [\tilde{\mathbf{p}}_n^{(i)}]_{1,k} > X_{\text{MAX}},\\
         & [\tilde{\mathbf{p}}_n^{(i)}]_{1,k} , \text{otherwise}.
    \end{aligned}
    \right.    \label{bound_particle}
\end{equation}
\begin{equation}
    [B\{\tilde{\mathbf{p}}_n^{(i)}\}]_{2,k}=
    \left\{  
    \begin{aligned}
         & Y_{\text{MIN}} , \text{if}\; [\tilde{\mathbf{p}}_n^{(i)}]_{2,k} < Y_{\text{MIN}},\\
         & Y_{\text{MAX}} , \text{if}\; [\tilde{\mathbf{p}}_n^{(i)}]_{2,k} > Y_{\text{MAX}},\\
         & [\tilde{\mathbf{p}}_n^{(i)}]_{2,k} , \text{otherwise}.
    \end{aligned}
    \right.    \label{bound_particle_y}
\end{equation}

In order to satisfy the constraint (\ref{distance_constraint}),  the fitness function is evaluated only for particles whose positions comply with the constraint. Consequently, the particle with antenna positions that violate constraint (\ref{distance_constraint}) would not impact the final result.

In each iteration of the PSO-based algorithm, the beamforming parameters are optimized prior to evaluating the fitness of the particles. This ensures the fitness reflects the locally optimal performance corresponding to each particle’s antenna positions. The PSO-based algorithm is summarized as \textbf{Algorithm \ref{PSObased_algorithm}}.
\begin{algorithm}[t]
    \renewcommand{\algorithmicrequire}{\textbf{Initialization:}}
	\renewcommand{\algorithmicensure}{\textbf{Output:}}
    \caption{PSO-based algorithm for optimization}
    \label{PSObased_algorithm}
    \begin{algorithmic}[1]
        \REQUIRE randomly generate $N_p$ particles with appropriate antenna positions, set $\mathbf{v}_J^{(0)}=0$.
        \FOR{$I =1 \cdots I_p$}
        \FOR{$J =1 \cdots N_p$}
            \STATE Update $\mathbf{v}_J^{(I)}$ as (\ref{velocity_update}).
            \STATE Update $\tilde{\mathbf{p}}_J^{(I)}$ as (\ref{positon_update}) and project as (\ref{bound_particle}).
            \IF{Position of particle satisfy the constraint (\ref{distance_constraint})}
            \STATE update $\mathbf{F},\mathbf{w}_r,\mathbf{w}_s,\mathbf{f}_{\text{UL}}$ and $\tilde{\mathbf{p}}_J^{(I)}$ as \textbf{Algorithm \ref{AObased_algorithm}}.
            \STATE Calculate $R_J^{(I)}$ as (\ref{bound_particle}) with $\tilde{\mathbf{p}}_J^{(I)}$.
            \IF{$R_J^{(I)}>R_{l}$}
                \STATE Set $\tilde{\mathbf{p}}_{l,J}=\tilde{\mathbf{p}}_J^{(I)}$ and $R_{l,J}=R_J^{(I)}$.
            \ENDIF

            \IF {$R_J^{(I)}>R_{g}$}
                \STATE Set $\tilde{\mathbf{p}}_{g}=\tilde{\mathbf{p}}_J^{(I)}$ and $R_{g}=R_J^{(I)}$.
            \ENDIF

            \ENDIF

            \ENDFOR

            \ENDFOR
        \STATE Set $\tilde{\mathbf{p}}=\tilde{\mathbf{p}}_{g}$ and optimize $\mathbf{F},\mathbf{w}_r,\mathbf{w}_s,\mathbf{f}_{\text{UL}}$ as \textbf{Algorithm \ref{param_all}}.           
        \ENSURE $\tilde{\mathbf{p}},\mathbf{F},\mathbf{w}_r,\mathbf{w}_s,\mathbf{f}_{\text{UL}}$.
        
    \end{algorithmic}
\end{algorithm}

\subsection{Convergence and Complexity Analysis}

The beamforming optimization process in \textbf{Algorithm \ref{param_all}} is guaranteed to be non-decreasing, as the parameters are iteratively updated to be optimal while adhering to the specified constraints. In \textbf{Algorithm \ref{AObased_algorithm}}, the antenna position is adjusted using the GA method, which ensures the non-decreasing nature of the objective function, and the beamforming matrices are refined as in \textbf{Algorithm \ref{param_all}}. In \textbf{Algorithm \ref{PSObased_algorithm}}, considering the global maximum fitness in the $i$-th iteration of the PSO process $R_g^{(i)}$, we have
\begin{equation}
    R_g^{(i+1)} \geq R_g^{(i)} .
\end{equation}
Consequently, the outcome is expected to be non-decreasing. Given the limitation of communication resources within the system, the objective function is bounded. Therefore, the convergence of all three algorithms is guaranteed.

For the sake of simplifying the complexity analysis, we assume that the number of transmit and receive antennas, $N_t$ and $N_r$, are of the same order of magnitude. Similarly, the number of downlink users $K_{\text{DL}}$, uplink users $K_{\text{UL}}$, and clutter sources C are assumed to be of the same order of magnitude. The computational complexities for updating $\mathbf{F}$ and $\mathbf{f}_{\text{UL}}$ are $\mathcal{O}(K_{\text{DL}} K_{\text{UL}} C N_t^2 + K_{\text{DL}} N_t^3)$ and $\mathcal{O}(K_{\text{DL}} K_{\text{UL}} N_r)$, respectively. The computation for updating $\mathbf{w}_r$ is $\mathcal{O}(K_{\text{UL}} N_r^3)$, while $\mathcal{O}(C N_r^3)$ refers to $\mathbf{w}_s$. Updating auxiliary variables $\boldsymbol{\mu},\boldsymbol{\xi}^{c,\text{DL}},\boldsymbol{\xi}^{c,\text{UL}},\boldsymbol{\xi}^s$ entails computational complexities of $\mathcal{O}( (K_{\text{UL}}^2+K_{\text{DL}} K_{\text{UL}}) N_t)$, $\mathcal{O}(K_{\text{DL}}^2 N_t)$, $\mathcal{O}(K_{\text{UL}} C N_t N_r )$, and $\mathcal{O}( C N_t N_r)$, respectively.
Hence, the computational complexity of \textbf{Algorithm \ref{param_all}} is $\mathcal{O}(I_b (K_{\text{DL}} K_{\text{UL}} C N_t^2 + K_{\text{DL}} N_t^3))$, which is dominated by the matrix inversion and other calculations in (\ref{precoding_update}), with $I_b$ being the iteration of \textbf{Algorithm \ref{param_all}}. Regarding the GA process for the antenna position, the computational complexity is $\mathcal{O}((I_{g,t}+ I_{g,r}) K_{\text{DL}} K_{\text{UL}} C N_t^2)$, where $I_{g,t}$ and $I_{g,r}$ correspond to the iterations of GA for the transmit and receive antennas, respectively. The overall computational complexity of \textbf{Algorithm \ref{AObased_algorithm}} is $\mathcal{O}(I_a((I_{g,t}+ I_{g,r}) K_{\text{DL}} K_{\text{UL}} C N_t^2 + I_b (K_{\text{DL}} K_{\text{UL}} C N_t^2 + K_{\text{DL}} N_t^3) ))$, with $I_a$ representing the iteration of AO. For the PSO-based algorithm, the complexity per iteration equals that of \textbf{Algorithm \ref{AObased_algorithm}}, thus the computational complexity of the algorithm is  $\mathcal{O}(I_p N_p (I_a((I_{g,t}+ I_{g,r}) K_{\text{DL}} K_{\text{UL}} C N_t^2 + I_b (K_{\text{DL}} K_{\text{UL}} C N_t^2 + K_{\text{DL}} N_t^3) )) )$.

\section{Simulation Result} \label{sec:4}
This section analyzes the proposed algorithm's performance through numerical simulations.

Uplink and downlink users, as well as clutters, are randomly situated around the BS. The elevation and azimuth AOAs and AODs are assumed to follow a uniform distribution, i.e., $\boldsymbol{\theta}_t, \boldsymbol{\theta}_r,\boldsymbol{\theta}_c, \boldsymbol{\phi}_t, \boldsymbol{\phi}_r, \boldsymbol{\phi}_c  \sim \mathcal{U}(0, \pi)$. 
The elevation and azimuth angles for the sensing target are $\theta_s=\pi/4$ and $\phi_s=0$.

 Distances for downlink users to the BS range from $40\,\text{m}$ to $70\,\text{m}$, while those for uplink users range from $30\,\text{m}$ to $60\,\text{m}$. The target-BS and clutter-BS distances are both between $20\,\text{m}$ and $40\,\text{m}$.
The complex RCS coefficients and the channel gain follow the standard complex Gaussian distribution, i.e., $\alpha_{s},\alpha_{c},\rho_{k,l} \sim \mathcal{C}\mathcal{N}(0,1)$. The free-space path losses for users, target, and clutters channels are denoted by $\eta=\big[\frac{\sqrt{G_l\lambda}}{4\pi d}\big]^2$ with $G_l=1$ since the BS uses omnidirectional antennas. 

Simulation parameters are detailed in Table \ref{table:param}, unless stated otherwise. Four schemes are compared in the simulation:

\begin{table}[t]
    \centering
    \caption{Simulation parameters}
    \begin{tabular}{|l<{\centering}|l<{\centering}|}
    \hline
        \label{table:param}
        \textbf{Parameter} & \textbf{Value} \\ \hline
        Transmit antenna number & $N_t=8$ \\ \hline
        Receive antenna number & $N_r=4$ \\ \hline     
        Uplink communication user number & $K_{\text{UL}}=3$ \\ \hline
        Downlink communication user number & $K_{\text{DL}}=3$ \\ \hline
        Clutter number & $C=3$ \\ \hline
        Channel path number & $L_p=10$ \\ \hline
        Carrier frequency & $f_c=30~\text{GHz}$ \\ \hline
        Wave length & $\lambda=0.01 \text{m}$ \\ \hline
        Self-interference distance & $r_{\text{SI}}=0.2~\text{m}$  \\ \hline
        Downlink transmit weight &  $\varpi_{c,\text{DL}}=0.3$ \\ \hline
        Uplink transmit weight &  $\varpi_{c,\text{UL}}=0.3$ \\ \hline
        Sensing weight & $\varpi_{s}=0.4$ \\ \hline
        Downlink transmit power & $P_{\text{DL}}= 30~\text{dBm}$ \\ \hline
        Uplink transmit power & $P_{\text{UL}}= 30~\text{dBm}$ \\ \hline
        Receive noise power & $\sigma_c^2=\sigma_s^2=-60~\text{dBm}$ \\ \hline
        Feasible range of movable antenna & \makecell{$X_{\text{MIN}}=Y_{\text{MIN}}=0$ \\$X_{\text{MAX}}=Y_{\text{MAX}}=0.06~\text{m}$}\\ \hline
        Minimum range between antennas & $D_0=0.005~\text{m}$ \\ \hline
        Random initial set number & $N_{RI}=300$ \\ \hline
        Particle number & $N_p=100$ \\ \hline
        PSO iteration & $I_p=50$ \\ \hline
    
    \end{tabular}
    
\end{table}

1) \textbf{FPA}: Both the transmit and receive antennas are fixed in the feasible region ranging $\lambda/2$ with the neighboring antenna, the beamforming matrices are updated as \textbf{Algorithm \ref{param_all}}.

2)\textbf{AO-MA}: Antennas are initially distributed uniformly in the feasible region, beamforming matrices and antenna positions are updated as \textbf{Algorithm \ref{AObased_algorithm}.}

3)\textbf{RI-MA}: Randomly generate $N_{RI}$ sets of initial antenna positions. Apply \textbf{Algorithm \ref{AObased_algorithm}} to optimize each of these sets and select the best result among all the optimized sets. 

4)\textbf{PSO-MA}: Randomly generate $N_p$ sets of antenna positions as particles, then optimize as \textbf{Algorithm \ref{PSObased_algorithm}}.

\begin{figure}[t]
    \centering
    \includegraphics[width=0.8\linewidth]{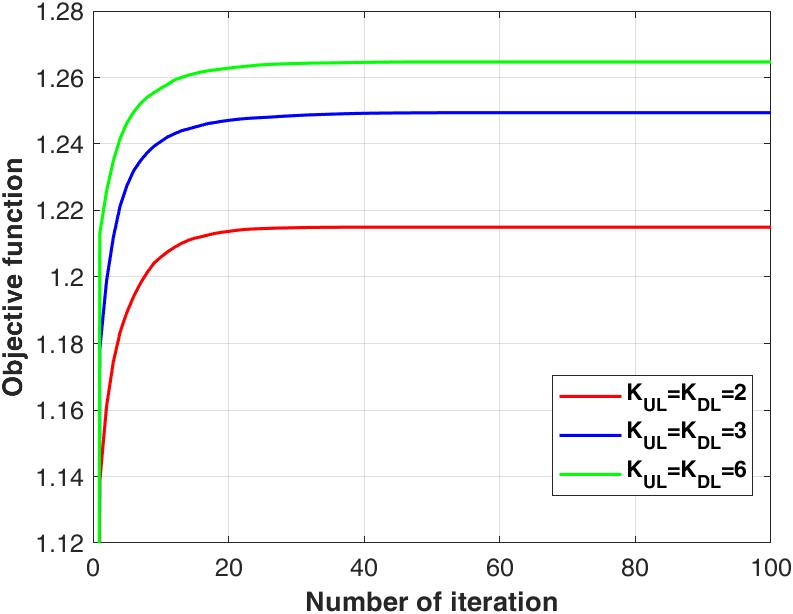}
    \caption{Convergence of Algorithm \ref{AObased_algorithm} with different number of users.}
    \vspace{-0.2cm}
    \label{fig:MA_converge}
\end{figure}

    Fig.\ref{fig:MA_converge} illustrates the convergence of \textbf{Algorithm \ref{AObased_algorithm}}. It is evident that, for varying user counts, the objective function rises with each iteration and converges after a few iterations, specifically around 30 iterations. 
    
\begin{figure}[t]
	\begin{minipage}{0.48\linewidth}
		\vspace{3pt}
		\centerline{\includegraphics[width=\textwidth]{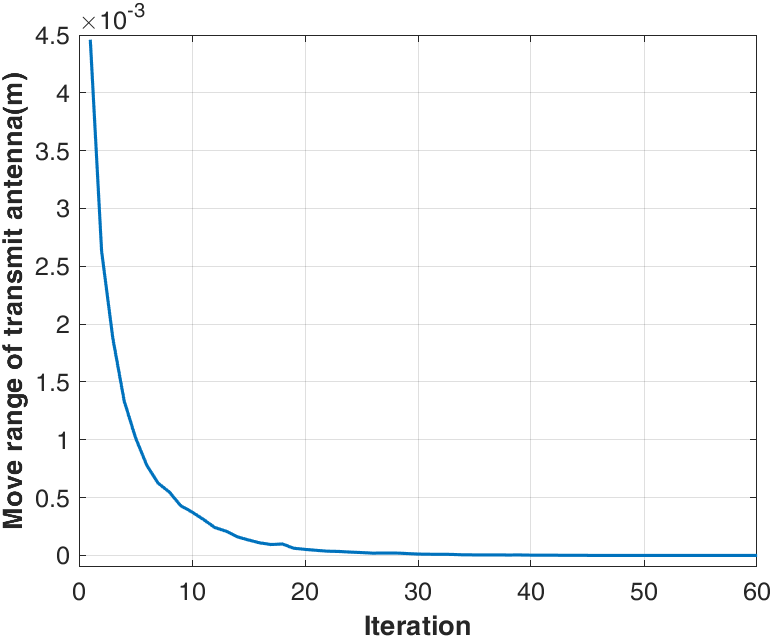}}
	 
		\centerline{(a) Transmit antennas}
	\end{minipage}
	\begin{minipage}{0.48\linewidth}
		\vspace{3pt}
		\centerline{\includegraphics[width=\textwidth]{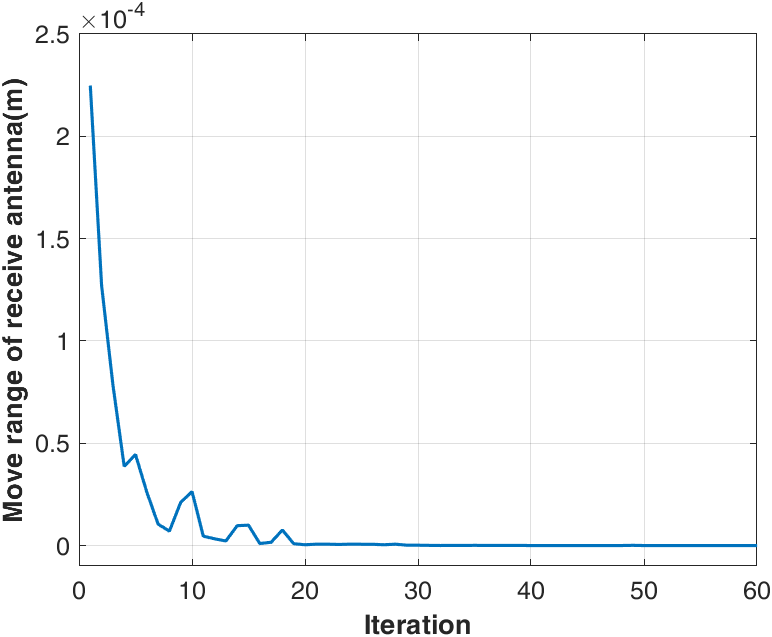}}
	 
		\centerline{(b) Receive antennas}
        \label{fig:rxr}
	\end{minipage}
 
	\caption{Total antenna moving distance in each iteration of Algorithm \ref{AObased_algorithm}.}
	\label{fig:txrx}
\end{figure}

    Fig.\ref{fig:txrx}  presents the total moving distance of all the transmit antennas and receive antennas in each iteration of \textbf{Algorithm \ref{AObased_algorithm}}, respectively. It can be observed that the moving distance exhibits a decreasing trend as iterations progress. Furthermore, the movement of the antenna stops at approximately 30 iterations, which aligns closely with the convergence of the objective function.  

     Notably, the maximum movement per iteration for the transmit antenna is only $0.0045~\text{m}$, equivalent to $0.45~\lambda$. And the total displacement of the transmit antenna is only roughly $0.015 \text{m}$, implying that the transmit antennas could only move in a confined region and cannot fully exploit the feasible region. 
     Consequently, it is unable to take full advantage of the movable antenna.
    For the receive antenna, the moving distance got even smaller. The maximum moving distance per iteration is less than $0.025~\lambda$ and the total displacement is less than $0.07~\lambda$.   

\begin{figure}[t]
    \centering
    \includegraphics[width=0.8\linewidth]{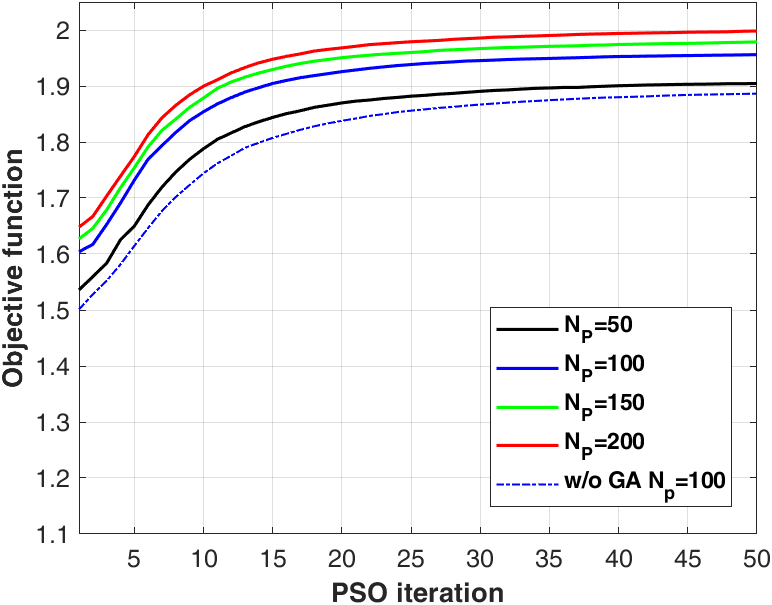}
    \caption{Objective function with different number of particles and iterations in Algorithm \ref{PSObased_algorithm}.}
    \vspace{-0.2cm}
    \label{fig:PSO}
\end{figure}

Fig.\ref{fig:PSO} shows the convergence of \textbf{Algorithm \ref{PSObased_algorithm}} with different particle numbers and iterations.
Results indicate that the objective function grows with both iterations and the number of particles. Moreover, it is important to note that the GA aids in locating a more optimal antenna position during the PSO process, as it can reach improved performance with fewer particles and iterations.

\begin{figure}[t]
    \centering
    \includegraphics[width=0.8\linewidth]{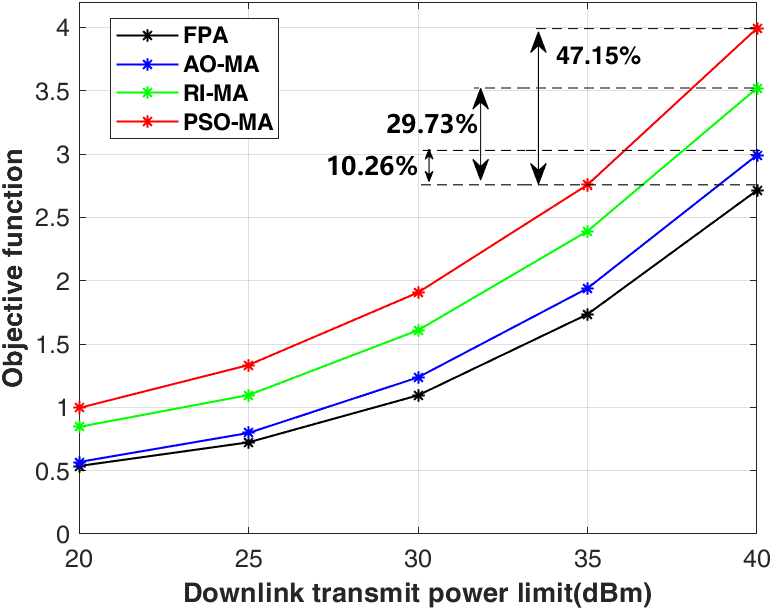}
    \caption{ISAC performance with different downlink transmit power.}
    \vspace{-0.2cm}
    \label{fig:downlinkpower}
\end{figure}

Fig.\ref{fig:downlinkpower} illustrates the ISAC performance relative to varying downlink transmit power, ranging from 20 dBm to 40 dBm. 
It indicates that movable antenna could enhance ISAC performance with different power levels.
When downlink transmit power $P_{\text{DL}}=40~\text{dBm}$, the objective function increased by 10.26\% with \textbf{AO-MA} compared to \textbf{FPA}. And there is a 29.73\% increment with \textbf{RI-MA}. For \textbf{PSO-MA}, the increment comes to 47.15\%. 
There is a larger gain when using \textbf{RI-MA} and \textbf{PSO-MA} , since they exploit a better sub-optimal antenna position, while \textbf{AO-MA} could only search for a local optimal result with its initial antenna positions.

\begin{figure}[t]
    \centering
    \includegraphics[width=0.8\linewidth]{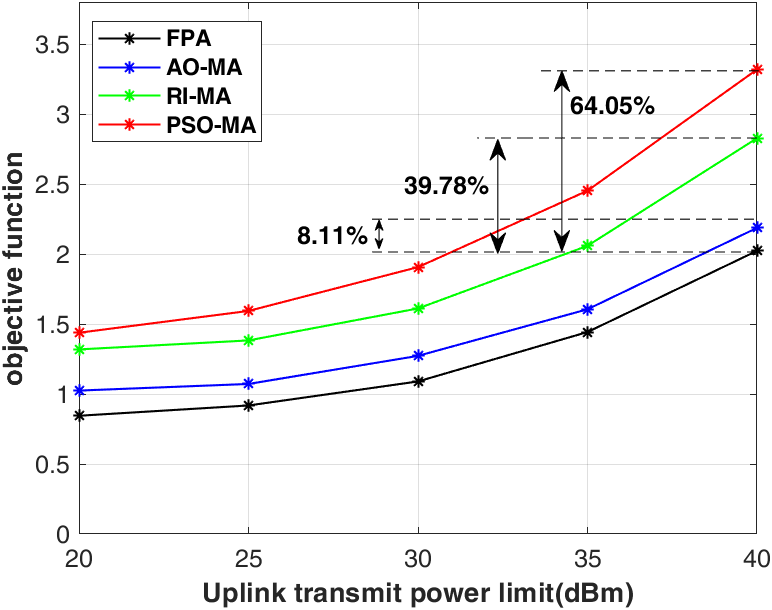}
    \caption{ISAC performance with different uplink transmit power.}
    \vspace{-0.2cm}
    \label{fig:uplinkpower}
\end{figure}

Fig.\ref{fig:uplinkpower} demonstrates the ISAC performance as uplink transmit power ranges from 20 dBm to 40 dBm. Analogous to downlink transmit power, the objective function increases with greater power. Across all simulated power levels, deploying movable antennas enhances system performance, with \textbf{PSO-MA} surpassing other methods due to its effective search for sub-optimal antenna positions.
At an uplink transmit power of $P_{\text{UL}}=40~\text{dBm}$, the gain over \textbf{FPA} is 8.11\% with \textbf{AO-MA}, 39.78\% with \textbf{RI-MA}, and 64.05\% with \textbf{PSO-MA}. Additionally, the performance enhancement from uplink power is less significant compared to downlink power for the same power increase, primarily because uplink power affects only communication, whereas downlink power influences both communication and sensing, which hold more weight in the objective function.

\begin{figure}[t]
    \centering
    \includegraphics[width=0.8\linewidth]{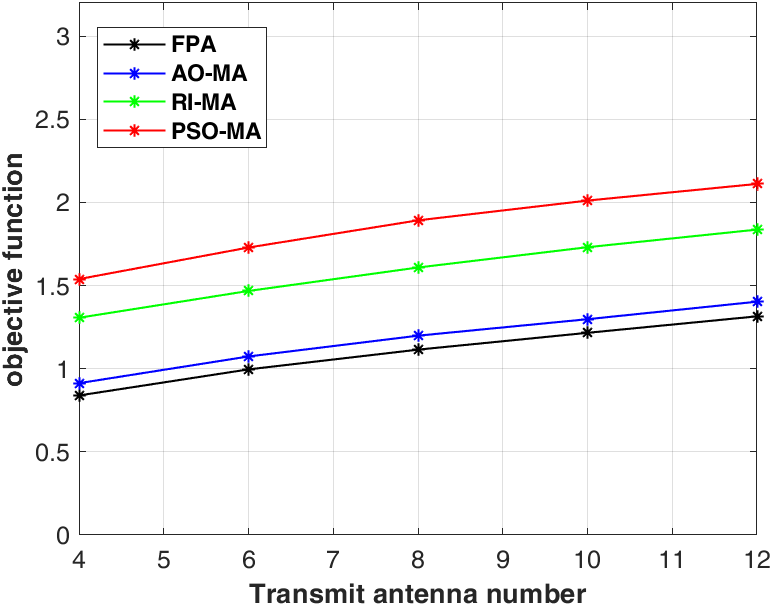}
    \caption{ISAC performance with different number of transmit antenna.}
    \vspace{-0.2cm}
    \label{fig:tx}
\end{figure}

\begin{figure}[t]
    \centering
    \includegraphics[width=0.8\linewidth]{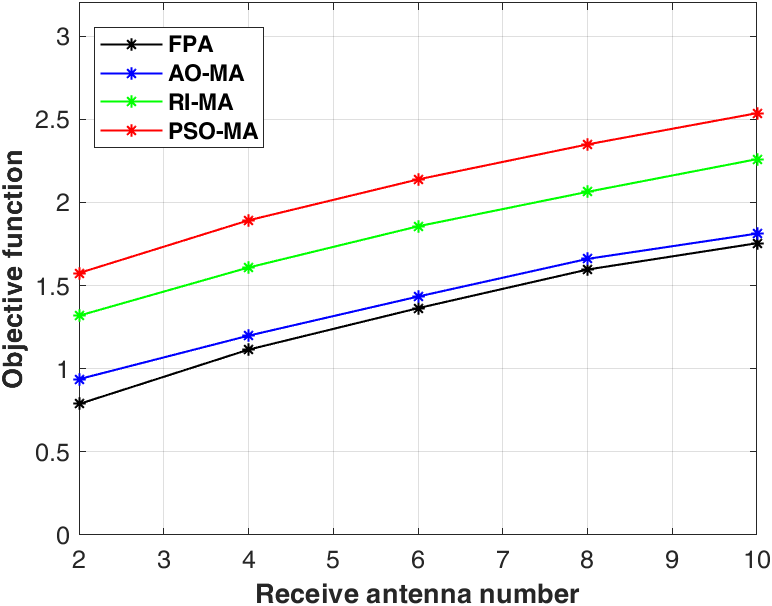}
    \caption{ISAC performance with different number of receive antenna.}
    \vspace{-0.2cm}
    \label{fig:rx}
\end{figure}

Fig.\ref{fig:tx} and Fig.\ref{fig:rx} show the performance variation with different transmit and receive antenna numbers. Performance improves across all methods as antenna numbers increase, with both exhibiting approximately linear growth.

\begin{figure}[t]
    \centering
    \includegraphics[width=0.8\linewidth]{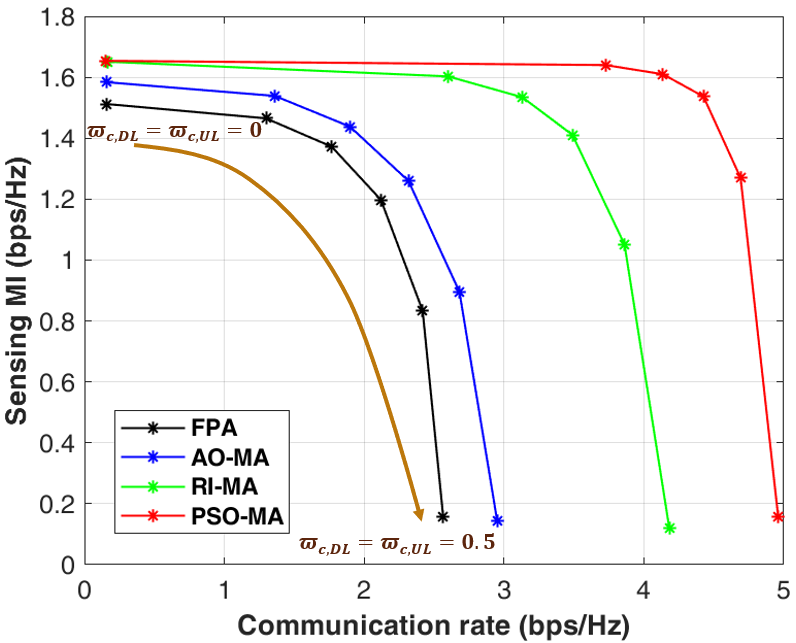}
    \caption{Communication rate and sensing MI with different weight.}
    \vspace{-0.2cm}
    \label{fig:wcwcul}
\end{figure}

In Fig.\ref{fig:wcwcul}, the trade-off between communication and sensing is depicted, where the weights for both uplink and downlink communication range from 0 to 0.5, and the sensing weight is adjusted such that their sum equals 1. Our analysis indicates that \textbf{RI-MA} and \textbf{PSO-MA} outperform \textbf{AO-MA} and \textbf{FPA} in various trade-off conditions. 
Notably, \textbf{PSO-MA} surpasses \textbf{RI-MA} in communication rate, though sensing MI sees only marginal enhancement. This is attributed to the ease of finding the optimal antenna position for sensing due to the single-path nature of the sensing channel, as opposed to the multi-path complexity of the communication channel. Consequently, algorithms that excel in optimizing antenna positions tend to have superior communication rates.

\begin{figure}[t]
    \centering
    \includegraphics[width=0.8\linewidth]{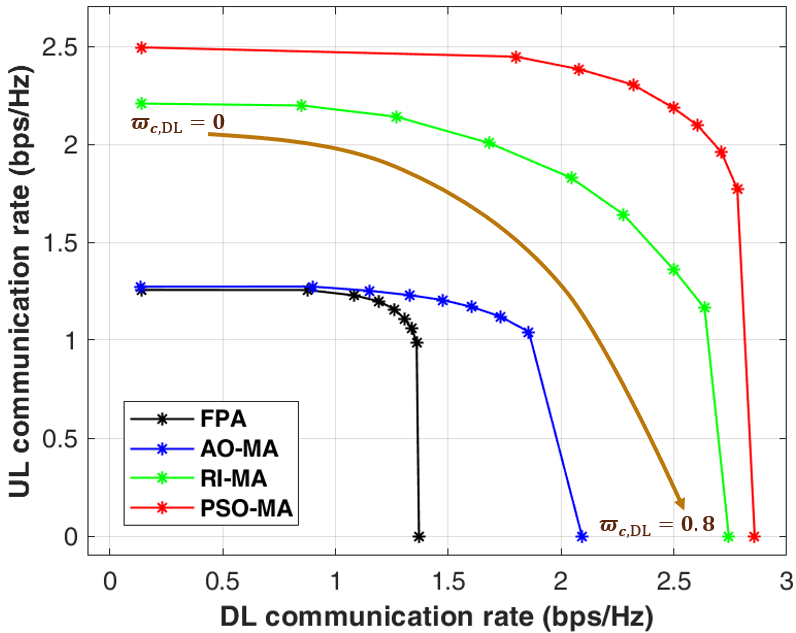}
    \caption{Uplink and downlink communication rate with different weight.}
    \vspace{-0.2cm}
    \label{fig:wc}
\end{figure}

Fig.\ref{fig:wc} illustrates the trade-off between downlink and uplink communication, setting $\varpi_{s}=0.2$ and varying $\varpi_{c,\text{DL}}$ from 0 to 0.8, with $\varpi_{c,\text{UL}}$ adjusted to keep their total at 1.
 Across all simulations, \textbf{AO-MA} shows only a slight improvement in uplink rate compared to \textbf{FPA} due to minimal receive antenna movement in the AO process, as seen in Fig.\ref{fig:txrx}. The close results stem from the similarity in receive antenna positions. Conversely, \textbf{RI-MA} and \textbf{PSO-MA} explore more positions, resulting in a notable increase in uplink rate. 
When the uplink weight is 0, transmission merely causes interference for sensing, and uplink communication is omitted to maximize the objective function.

\begin{figure}[t]
    \centering
    \includegraphics[width=0.8\linewidth]{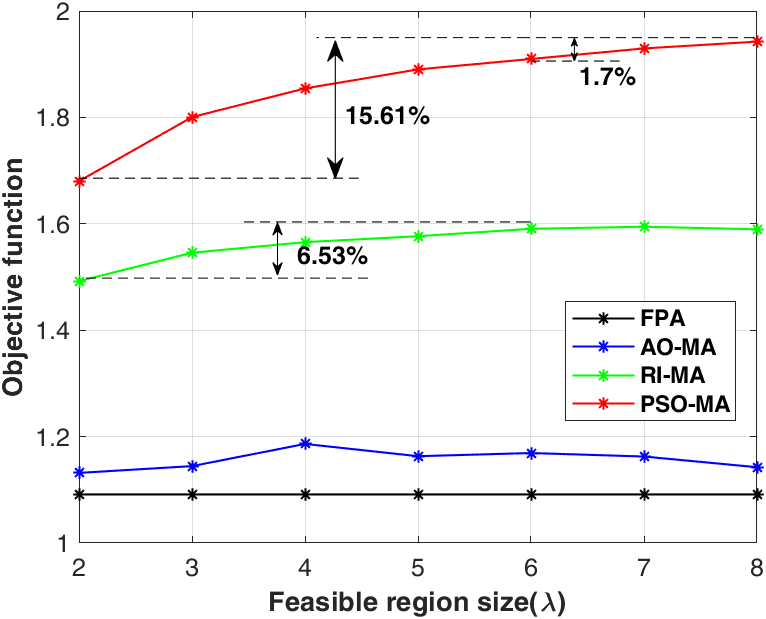}
    \caption{ISAC performance with different feasible region size.}
    \vspace{-0.2cm}
    \label{fig:range}
\end{figure}

Fig.\ref{fig:range} illustrates performance across various feasible region sizes. The performance of \textbf{FPA} remains consistent regardless of region size, as the antennas do not benefit from the movable region.
The performance of \textbf{AO-MA} fluctuates with no clear improving trend. The reason is that the antennas in \textbf{AO-MA} are uniformly distributed in the  feasible region and thus the initial antenna positions alter with feasible region size. And the \textbf{AO-MA} could only get a local optimal antenna position from its initial position, which could not provide a higher gain with a larger feasible region. 



Unlike \textbf{AO-MA}, both \textbf{RI-MA} and \textbf{PSO-MA} could search for sub-optimal antenna positions with various initial antenna positions, showing growth with an enlarged feasible region.

However, due to the limited number of random initial positions in \textbf{RI-MA}, the method becomes inefficient in searching for optimal results within extensive regions, where numerous potential positions exist.
Results indicate a 6.53\% increase in the objective function when the region grows from 2$\lambda$ to 6$\lambda$, with minimal gain beyond 6$\lambda$.
In contrast, \textbf{PSO-MA} achieves a 15.6\% increase as the region extends from 2$\lambda$ to 8$\lambda$, and still achieves a 1.7\% rise between 6$\lambda$ and 8$\lambda$. This demonstrates that \textbf{PSO-MA} harnesses the feasible region more effectively by efficiently locating sub-optimal antenna positions.

\section{Conclusion} \label{sec:5}

In this paper, we discussed a communication FD mono-static sensing integrated ISAC system using MA. The system supports multiple users for uplink and downlink communication and simultaneously senses a target while managing interference and SI between transmit and receive antennas. An optimization problem is formulated to enhance performance across communication and sensing functionalities by optimizing beamforming matrices, uplink power, and antenna positions. Due to the problem's non-convex nature, an AO-based algorithm is proposed along with a GA method to find a local optimal solution,
and a PSO-based algorithm is also proposed to better explore the feasible region. Simulations show that the AO-based method enhances performance compared to fixed antenna system, and the PSO method further improves outcomes by effectively utilizing the flexibility of movable antennas. The PSO-based algorithm is also adept at handling more complex channels and larger feasible regions.



\bibliography{IEEEabrv,refs}
\bibliographystyle{IEEEtran}


\end{document}